\documentclass[10pt,a4paper]{article}

\usepackage[utf8]{inputenc}
\usepackage[T1]{fontenc}
\usepackage{lmodern} 

\usepackage[a4paper,top=3cm,bottom=3cm,left=1.8cm,right=1.8cm]{geometry}
\usepackage{amsmath, amsfonts, amssymb, amsthm}
\numberwithin{equation}{section}

\usepackage{graphicx}
\usepackage{subcaption}
\usepackage{tikz}
\usetikzlibrary{arrows.meta, patterns, decorations.pathmorphing}

\usepackage{enumitem}

\usepackage{hyperref}
\usepackage[numbers,sort&compress]{natbib}
\usepackage{fancyhdr}

\newcommand{\shorttitle}{Tailoring dispersion and evanescent modes}
\newcommand{\shortauthor}{L. Rouhi et al.}

\usepackage[final]{changes} 
\usepackage{soul}

\soulregister\ref7
\soulregister\subref7
\soulregister\eqref7
\soulregister\citep7
\soulregister\pageref7

\definecolor{RAcolor}{RGB}{220,230,255} 
\definecolor{RBcolor}{RGB}{255,220,220} 
\definecolor{Anothercolor}{RGB}{252,250,167}

\definechangesauthor[color = RAcolor, name={Reviewer A}]{RA}
\definechangesauthor[color = RBcolor, name={Reviewer B}]{RB}
\definechangesauthor[color = Anothercolor, name={Other}]{Oth}

\setauthormarkup{}
\setaddedmarkup{{\sethlcolor{authorcolor}\textcolor{black}{\hl{#1}}}}

\date{}

\begin{document}

	\title{Tailoring dispersion and evanescent modes in multimodal nonlocal lattices using positive-only interactions}

	\author{%
		\begin{tabular}{c}
		\textbf{Lucas Rouhi}$^{1}$ \quad \textbf{Christophe Droz}$^{1}$ \\[0.6em]
		\small $^{1}$Univ.~Gustave Eiffel, Inria, COSYS--SII, I4S, Rennes, France \\[0.3em]
		\small \texttt{lucas.rouhi@inria.fr}
		\end{tabular}
	}

	\maketitle
	\begingroup
	\small 
	\noindent\textbf{Highlights}
	\begin{itemize}[leftmargin=1.5em]
    	\item Customization of band diagrams through nonlocal interactions is key to advanced phononic design.
        \item Versatility of the method enables selective customization of Rotons, ZGV and GVD indicators.
        \item The method applies to a broad range of lattices with nonlocal spring and beam interactions.
    	\item We achieved passive and stable mechanical systems through real, positive-only nonlocal stiffnesses.
    	\item We control multimodal evanescent decay inside bandgaps by tuning the complex wavenumbers.
	\end{itemize}
	\endgroup
	
	\begin{abstract}
	\setlength{\parindent}{0pt}
	Metamaterials derive their unconventional properties from engineered microstructures,
	with periodic lattices providing a versatile framework for modeling wave propagation.
	Dispersion relations, obtained from Bloch-Floquet theory, govern how waves propagate, attenuate, or localize within such systems.
	Extending interactions beyond nearest neighbors, through nonlocality, substantially enriches the design space of band diagrams,
	enabling phenomena such as negative or zero group velocities, roton-like extrema, and band-gap localization.
	However, existing approaches to dispersion tailoring often rely on analytical formulations or Fourier-based identifications,
	which become impractical for complex coupling mechanisms and offer limited control over physical constraints such as stiffness positivity.

	This work introduces a general interpolation-based framework for customizing dispersion relations in uniform nonlocal lattices.
	Rather than reconstructing full dispersion curves, the method enforces prescribed frequency-wavenumber points as interpolation constraints,
	enabling localized and tunable control of wave behavior.
	The formulation is applied to both spring- and beam-interaction lattices, and demonstrated on an Euler-Bernoulli beam model with adjustable nonlocal couplings.
	Through systematic parameter tuning, the framework enables the creation of rotons,
	the adjustment of group-velocity dispersion, and the design of evanescent waves with controlled exponential decay within band gaps,
	all while ensuring real, positive-only stiffness parameters and passive mechanical behavior.
	
	Altogether, this parametric interpolation strategy provides a physically consistent and computationally
	efficient route for engineering advanced phononic functionalities in periodic nonlocal systems.
	\end{abstract}


	\section{Introduction}
	Metamaterials are artificially engineered composites whose microstructural design grants access to physical properties absent in conventional materials.  
	Among them, \added[id = Oth]{phononic material} stand out for their ability to manipulate mechanical waves through carefully designed periodic architectures.  
	Such structural periodicity enables tailored control of wave propagation, with key applications in vibration damping, noise mitigation, and acoustic cloaking.  
	As a result, metamaterials play an essential role across diverse industries, ranging from civil engineering and automotive design to aerospace and acoustic insulation systems.

	A common mathematical abstraction for modeling such materials is the \emph{periodic lattice}, obtained by infinitely repeating an elementary unit cell.  
	These unit cells may range from simple monoatomic arrangements to more elaborate structures such as \added[id = Oth]{diatomic lattices or multi-mass structures.}  
	Exploiting periodicity,
	the dynamics of the infinite lattice reduce to supercells governed by setsof ordinary differential equations~\citep{brillouin1946wave,hussein2014dynamics},
	providing an analytically tractable yet powerful framework for investigating wave phenomena.

	At the heart of this analysis lies the Bloch-Floquet theory, which expresses solutions as harmonic waves modulated by spatially periodic functions.  
	This formalism naturally yields dispersion relations connecting angular frequency and wavenumber.  
	Dispersion curves, in turn, are indispensable tools for identifying propagating, evanescent, and attenuating modes,
	thereby offering deep insight into transmission and localization mechanisms.

	While conventional models typically assume nearest-neighbor couplings,
	nonlocal lattice models have attracted increasing attention since Brillouin's pioneering work in 1946~\citep{brillouin1946wave},
	whose archetypal example, a monoatomic chain with long-range spring interactions, is illustrated in Figure~\ref{graph:unif_monoatom}.
	His original formulation of nonlocal lattices has been recently revived and extended in various modern contexts,
	highlighting their potential for advanced control of wave propagation.
	By extending interactions beyond nearest neighbors,
	nonlocality introduces additional tunable parameters and thereby greatly enriches the design space of dispersion relations.  
	This expanded framework allows access to remarkable wave phenomena, including
	negative or hypersonic group velocities~\citep{carcaterra2019long,rezaei2021wave},
	roton-like dispersion profiles~\citep{iglesias2021experimental,wang2022nonlocal,zhu2022observation,iorio2024roton},
	dirac cones~\cite{paul2025nonlocality},
	frozen evanescent modes~\citep{chen2024anomalous,wang2025roton},
	metadamping~\citep{banerjee2024enhanced}
	and the emergence of additional transient modes~\citep{brillouin1946wave,kazemi2023non}.  
	\added[id = RB]{Although nonlocal lattices are not the only class of metamaterials exhibiting such phenomena,
	their parametric versatility offers a particularly promising strategy for tailoring wave behavior in periodic lattices.}

	\begin{figure}[ht]
    	\centering
    	\begin{tikzpicture}[scale=1.2, every node/.style={scale=1.2}]
        	\foreach \x in {0,1,2,3,4,5} {
            	\node[draw, circle, fill=black] (atom\x) at (\x, 0) {};
        	}

        	\foreach \x in {0,1,2,3,4} {
            	\draw[thick] (atom\x) -- (atom\the\numexpr\x+1\relax);
        	}

        	\foreach \x in {0,1,2} {
            	\draw[thick, color=blue] (\x*2, 0.16) -- (\x*2, 0.6);
            	\draw[thick, color=blue] (\x*2+1, -0.16) -- (\x*2+1, -0.6);
       		}
        	\draw[thick, color=blue] (0, 0.6) -- (5, 0.6);
        	\draw[thick, color=blue] (0, -0.6) -- (5, -0.6);

        	\node at (1, 0.35) {$m$};
        	\draw[<->] (3.25,-0.15) -- (3.75,-0.15) node[midway, below] {$l$};
        	\node at (1.5, 0.25) {$\beta_1$};
        	\node at (4.25, 0.35) {$\beta_2$};
        	\node at (0.75, -0.35) {$\beta_2$};

        	\draw[dotted, thick] (atom0) -- (-0.7, 0);
        	\draw[dotted, thick, color=blue] (0,0.6) -- (-0.7, 0.6);
        	\draw[dotted, thick, color=blue] (0,-0.6) -- (-0.7, -0.6);
        	\draw[dotted, thick] (atom5) -- (5.7, 0);
        	\draw[dotted, thick, color=blue] (5,0.6) -- (5.7, 0.6);
        	\draw[dotted, thick, color=blue] (5,-0.6) -- (5.7, -0.6);
    	\end{tikzpicture}
    	\caption{Monoatomic chain with non-local \added[id = Oth]{spring} interactions up to order $P=2$. Local interactions are shown in black, while higher-order couplings are highlighted in blue.
		\added[id = Oth]{The stiffness coefficients $\beta_1$ and $\beta_2$ correspond to first- and second-order springs, respectively.
		The distance between adjacent unit cells is denoted by $l$.}}
    	\label{graph:unif_monoatom}
	\end{figure}

	The design of nonlocal lattices can be understood as a two-step process: selecting a geometry of nonlocality to be periodically repeated,
	and specifying the physical nature of the interactions.  
	Several nonlocal lattice geometries have been explored in the literature.  
	The most common configuration connects each node to its $P$-th neighbors~\citep{edge2025discrete, kazemi2023non, guarracino2024local, chen2023phonon, edge2025spring},
	while other works have considered diatomic extensions introducing two distinct species of nodes~\citep{kazemi2023non, ongaro2025closed},
	or nonlocal architectures incorporating local resonators~\citep{ghavanloo2019wave}.  
	Different physical realizations of the coupling mechanisms have also been investigated.  
	Most studies employ spring-based interactions~\citep{kazemi2023non, guarracino2024local, chen2023phonon, edge2025spring},
	but alternative implementations have been proposed, including networks of coaxial cables~\citep{chen2023nonlocal},
	reconfigurable assemblies using model-construction kits such as Meccano\textsuperscript{TM}~\citep{chaplain2023reconfigurable},
	electromagnetic forces~\citep{rezaei2021wave},
	and Euler-Bernoulli beam interactions~\citep{edge2025discrete}.  

	The central challenge lies in precisely shaping dispersion relations.  
	Brillouin~\citep{brillouin1946wave} already recognized that the dispersion relation of a generalized nonlocal spring lattice could be expressed as a Fourier series,
	with spring stiffnesses corresponding, up to scaling, to Fourier coefficients.  
	In principle, this identification allows any target dispersion curve to be approximated by assigning stiffness values according to Fourier decomposition.  
	However, not all dispersion profiles are physically realizable.  
	Kazemi \textit{et al.}~\citep{kazemi2023non} clarified the feasibility conditions of this approach,
	showing that while certain constraints exist, such as linear behavior at low wavenumbers and vanishing group velocity at high wavenumbers, they remain relatively mild
	\added[id = Oth]{thereby allowing for a wide range of admissible dispersion curve designs.}

	Subsequent extensions included the generalization of Brillouin’s method to diatomic lattices with spring-based nonlocality by \added[id = Oth]{Kazmi \textit{et al.}}~\citep{kazemi2023non}
	as well as the formulation of dual nonlocal interactions by \added[id = Oth]{Paul \textit{et al.}}~\citep{paul2024complete}.
	More recently, Edge \textit{et al.}~\citep{edge2025spring} proposed an initial approach for parameterizing evanescent modes in nonlocal spring-interaction models.
	Yet, these strategies remain difficult to apply when the coupling mechanism departs from simple springs.
	They rely on closed-form analytical expressions of the dispersion relation, often unavailable in practice,
	and on linear mappings between stiffnesses and Fourier coefficients that break down for more complex interaction laws,
	leading instead to nonlinear identification problems.

	\added[id = RA]{In addition, nothing in the general identification process guarantees that the resulting stiffness coefficients remain positive.
	Although negative-stiffness elements have become an active and productive research topic in active metamaterial systems, with substantial progress reported in recent years \citep{tan2025negative},
	parameterization objectives may also arise for purely passive nonlocal models, i.e., without energy injection into the system.}
	Edge \textit{et al.}~\citep{edge2025spring} proposed sufficient criteria on admissible target dispersion curves to ensure positive stiffnesses.
	\added[id = RB]{However, this criterion takes the form of an analytical constraint on the dispersion curve that is difficult to verify
	in practice and is only applicable to the monoatomic spring-interaction model.}
	This limitation stems from the very objective of reconstructing an entire target dispersion curve, which tends to overconstrain the design problem.
	In many practical contexts, however, the interest lies not in reproducing a full curve but in tailoring specific points
	or regions that govern distinctive wave phenomena,
	such as rotons~\citep{iglesias2021experimental,wang2022nonlocal,zhu2022observation,iorio2024roton},
	frozen evanescent modes~\citep{chen2024anomalous,wang2025roton}, or Dirac cones~\citep{paul2025nonlocality},
	or in locally shaping the dispersion to control group-velocity dispersion or band-gap properties~\citep{edge2025spring}.

	Motivated by these limitations, the present work introduces a complementary paradigm for dispersion tailoring.
	Rather than reconstructing an entire target curve, the proposed approach focuses on localized customization:
	specific frequency-wavenumber points are prescribed as interpolation constraints that the dispersion relation must satisfy.
	\added[id = RB]{This formulation enables direct and flexible control over selected regions of the spectrum,
	allowing both global and local parameterization of the model dispersion curve.}
	The general framework is first established and then applied to the Euler-Bernoulli beam lattice model introduced by Edge \textit{et al.}~\citep{edge2025discrete},
	which is here reformulated to incorporate tunable nonlocal couplings and multiple degrees of freedom per node.
	The resulting multimodal dynamics allow simultaneous control of several interacting wave branches within a unified setting.
	Several representative applications are presented to illustrate the method's capabilities, including the creation of rotons,
	the tuning of group-velocity dispersion, and the controlled design of evanescent modes within band gaps.
	In addition, the interpolation scheme naturally accommodates design constraints, such as enforcing positive stiffness parameters,
	to ensure passive and mechanically stable interactions at all stages of the tuning process.
	Altogether, the framework provides a unified and physically consistent foundation for engineering wave phenomena in nonlocal lattices:
	it applies seamlessly to both spring and beam systems, guarantees stability through real positive stiffness coefficients,
	and enables fine control of propagation and attenuation mechanisms via complex dispersion analysis.

	\section{Tuning of dispersion relations in uniform nonlocal lattices}
	\label{sec:Tuning_UniformNonlocal}
	The ability to tailor the dispersion characteristics of nonlocal lattices systems comes from a simple yet fundamental observation:  
	each nonlocal interaction, regardless of its underlying physical nature, is governed by one or more tunable parameters.  
	Introducing additional nonlocal interactions effectively increases the number of degrees of freedom in the system,  
	which can in turn be exploited to shape the dispersion curves in a controlled manner.  
	
	The approach developed here follows a two-step rationale.  
	First, a formal description of the class of \emph{uniform nonlocal lattices} is introduced,
	providing a precise mathematical framework in which the dispersion relation can be expressed and analyzed.  
	Second, a systematic method is proposed for selecting the interaction parameters so as to enforce  
	prescribed points on the dispersion curves, thus achieving the desired wave–propagation properties.

		\subsection{Uniform nonlocal models}
		The lattices considered in this work are infinite, one-dimensional structures:  
		their nodes are aligned along a single axis, and waves propagate exclusively in this direction.  
		To introduce the concept of nonlocality, let the nodes be equally spaced at a distance $l$.  
		A reference node is denoted by $x_0$, and the next node along the positive direction is $x_1$, located at a distance $l$ from $x_0$.  
		This indexing extends such that $x_n$ denotes the node located at $nl$ from $x_0$ in the positive direction, and $x_{-n}$ in the opposite direction.
		
		A nonlocal interaction of order $p$ couples two nodes separated by a distance $pl$, that is, between $x_n$ and $x_{n+p}$.  
		A lattice is said to be nonlocal of order $P$ if interactions extend to at most $P$ neighbors.  
		In the present study, the focus is on \emph{uniform nonlocal interactions}, meaning that if an order-$p$ interaction exists between a given pair of nodes,
		then every pair of nodes separated by $pl$ is coupled by the same type of interaction, characterized by identical parameters.

		Once the interaction model between nodes is fixed, a uniform nonlocal lattice of order $P$ is entirely characterized by its maximum order $P$  
		and the set of parameters $\boldsymbol{\beta} = (\beta_p)$ configuring each interaction of order $p$.  
		These parameters are generally real and positive, but may acquire a imaginary part in the presence of damping.

		Each node possesses $d$ mechanical degrees of freedom, and the displacement of node $x_n$ from equilibrium is denoted by $\mathbf{u}_n \in \mathbb{R}^d$.  
		For a uniform nonlocal lattice of order $P$, the equations of motion for each $\mathbf{u}_n$ take the generic form
		\begin{equation}
			\mathbf{M}\,\partial_t^2\mathbf{u}_n + \mathbf{K}\,[\mathbf{u}_{n-P} \dots \mathbf{u}_n	 \dots \mathbf{u}_{n+P}	]^T = 0 
		\label{eq:motion_nonlocal_model}
		\end{equation}
		where $\mathbf{M}$ and $\mathbf{K}$ are matrices of size $d\times d$ and $d\times (2P+1)$, respectively.
		
		To analyze wave propagation in such systems, Floquet-Bloch theory is employed~\cite{hussein2014dynamics}, seeking solutions of the harmonic form 
		\begin{equation}
			\mathbf{u}_n(t) = \mathbf{\Phi}\lambda^n e^{i\omega t}
		\label{eq:Floquet_form}
		\end{equation}
		where $\lambda \in \mathbb{C}$ is the propagation constant, $\omega \in \mathbb{R}_+$ the angular frequency,
		and $\boldsymbol{\Phi} \in \mathbb{C}^d$ the mode shape.
		It is often convenient to set $\lambda = e^{-il\kappa}$, where $\kappa$ denotes the wavenumber.
		
		The detailed derivation is provided in Appendix~\ref{part:Dispersion_Relation}, substituting~\eqref{eq:Floquet_form} into~\eqref{eq:motion_nonlocal_model}
		leads to the dispersion relation  
		\begin{equation}
			\left[  \sum_{p=1}^P \mathcal{K}_p(\lambda, \beta_p) - \omega^2\mathbf{M} \right] \mathbf{\Phi} = 0
		\label{eq:nonlocal_unif_dispersion}
		\end{equation}
		with
		\begin{equation}
			\mathcal{K}_p(\lambda, \beta_p) = \mathbf{A}_{p}^+(\beta_p) +
			\mathbf{B}_{p}^+(\beta_p)\lambda^p + \mathbf{A}_{p}^-(\beta_p) + \mathbf{B}_{p}^-(\beta_p)\lambda^{-p}
		\end{equation}
		where $\mathbf{A}_{p}^+(\beta_p)$, $\mathbf{A}_{p}^-(\beta_p)$, $\mathbf{B}_{p}^+(\beta_p)$ and $\mathbf{B}_{p}^-(\beta_p)$
		are square matrix of size $d\times d$.
		For a fixed $\lambda \in \mathbb{C}$ and set of parameters $\boldsymbol{\beta}$, solving~\eqref{eq:nonlocal_unif_dispersion} yields 
		$d$ admissible angular frequencies $\omega_j(\lambda, \boldsymbol{\beta})$, with $1 \leqslant j \leqslant d$.
		
		This formal framework sets the stage for a systematic tuning strategy, 
		where the parameters $\beta_p$ are chosen so as to enforce specific target points on the dispersion curves.
		The procedure is detailed in the next subsection.

		\subsection{Tuning method}
		\label{sec:tuning}
		Building upon the above framework, the aim is to parametrically control  the dispersion curves of uniform nonlocal lattices.  
		In this context, ``parametric control'' means prescribing specific points $(\lambda_i, \omega_i)$
		that the dispersion relation of the model must satisfy.  
		Formally, for each prescribed $\lambda_i$, one of the $d$ angular frequencies  
		$\omega_j(\lambda_i, \boldsymbol{\beta})$ must satisfy
		\begin{equation}
    			\omega_j(\lambda_i, \boldsymbol{\beta}) = \omega_i.
		\end{equation}
		For the various parameterization objectives discussed in Section~\ref{part:Engineering_wave_phenomena},  
		it can be more convenient to operate directly on the dispersion curves,  
		and thus to prescribe target points in the form $(\kappa_i, \omega_i)$.
		The procedure begins by selecting the maximum nonlocal order $P$ and fixing the interaction model between the nodes.  
		The chosen model depends on $P$ physical parameters $\beta_p$ characterizing the nonlocal couplings.  
		It is then possible to select \added[id = Oth]{at most} $P$ target points $(\lambda_i, \omega_i)$ (or ($\kappa_i, \omega_i)$) to be interpolated by the dispersion relation.  
		This leads to the nonlinear system
		\begin{equation}
			    \omega_j(\lambda_i, \boldsymbol{\beta})^2 - \omega_i^2 = 0, \quad i \in \{1, \dots, P\},
		\label{eq:system_interpolation}
		\end{equation}
		with the $P$ unknowns $\boldsymbol{\beta} = (\beta_p)$.  
		\added[id = RB]{In general,~\eqref{eq:system_interpolation} is nonlinear with respect to $\boldsymbol{\beta}$,
		so the existence and uniqueness of a solution are not guaranteed.  
		Both properties depend on the specific nonlocal model under consideration.  
		For beam interaction models, situations involving non-existence of solutions as well as examples of non-uniqueness are discussed in Appendix~\ref{sec:other_dispersion_curves}.
		However, squaring the angular frequencies makes it possible, in certain cases, to obtain a linear system, for instance
		in the monoatomic chain with uniform nonlocal spring interactions.
		In this setting, as detailed in Appendix~\ref{part:monoatomic-lattice}, existence and uniqueness of the solution are guaranteed.}
		
		The applicability and versatility of the interpolation framework are best appreciated through concrete examples.  
		Two representative cases are considered in this work:  
		the fundamental monoatomic chain with uniform nonlocal spring interactions, presented in Appendix~\ref{part:monoatomic-lattice},  
		and the Euler-Bernoulli beam lattice,
		discussed in the next section, which serves to illustrate the method within a physically richer and structurally more complex context.

	\section{Application to an Euler-Bernoulli beam lattice}
	\label{section:Euler-BernoulliBeamLattices}
	In this model, each node possesses two degrees of freedom ($d=2$), corresponding to transverse displacement and rotation,
	and is connected to neighboring nodes by Euler-Bernoulli beams.  
	Each connecting beam is characterized by its Young's modulus $E_p$ and second moment of inertia $I_p$, defining the stiffness parameter
	\[
    	\beta_p = E_p I_p.
	\]

	The nearest-neighbor configuration of such lattices has been studied by Madine and Colquitt~\cite{madine2021dynamic},
	while uniform nonlocal extensions with identical beam properties for all interaction orders were introduced by Edge \textit{et al.}~\cite{edge2025discrete}.  
	The present formulation further generalizes this setting by allowing the mechanical properties of the connecting beams to depend explicitly on the interaction order $p$.

	Following the formalism of Section~\ref{sec:Tuning_UniformNonlocal} and as detailed in Section~\ref{sec:DispersionBeamLattices},
	the dispersion relation of order-$P$ lattices takes the form of Eq.~\eqref{eq:nonlocal_unif_dispersion}, with explicit coefficient matrices
		\begin{equation}
			\mathbf{A}_{p}^+(\beta) =
			\frac{\beta_p}{p^3}
			\begin{bmatrix}
				12 	&6p	\\
				6p	&4p^2
			\end{bmatrix}
			\text{  , }
			\mathbf{B}_{p}^+(\beta) =
			\frac{\beta_p}{p^3}
			\begin{bmatrix}
				-12 	&6p	\\
				-6p	&2p^2
			\end{bmatrix}
		\end{equation}
		\begin{equation}
			\mathbf{A}_{p}^-(\beta) =
			\frac{\beta_p}{p^3}
			\begin{bmatrix}
				12 	&-6p	\\
				-6p	&4p^2
			\end{bmatrix}
			\text{  , }
			\mathbf{B}_{p}^-(\beta) =
			\frac{\beta_p}{p^3}
			\begin{bmatrix}
				-12 	&-6p	\\
				6p	&2p^2
			\end{bmatrix}
		\end{equation}
		The corresponding lumped mass matrix is
		\begin{equation}
			\mathbf{M} =
			\begin{bmatrix}
				1 & 0		\\
				0 &\mu
			\end{bmatrix}
		\end{equation}
		with $0<\mu\ll 1$.
		For each fixed wavenumber, the generalized eigenvalue problem~\eqref{eq:nonlocal_unif_dispersion}
		admits two distinct angular frequencies.  
		In this work, only cases where $\Im(\omega) = 0$ are considered.
		The two positive roots of~\eqref{eq:nonlocal_unif_dispersion} are denoted by $\omega_- \leq \omega_+$.

		In Appendix~\ref{sec:AnalyticalDispersionBeam}, analytical expressions for $\omega_-$ and $\omega_+$ are fully derived,
		along with a discussion on how these closed-form results can be used to simplify the interpolation
		system~\eqref{eq:system_interpolation} for this model.  
		However, such analytical formulations are not required for the present approach,
		and numerical computations are employed throughout this work.

		The dispersion relation therefore consists of two branches.
		The interpolation method introduced earlier can be directly applied to this model.  
		Interpolation points $(\lambda_i, \omega_i)$ are selected with explicit specification of the branch at each point,
		leading to a system of the form~\eqref{eq:system_interpolation}.  
		Depending on the design objective, either a single branch or both branches may be interpolated simultaneously.

\begin{figure}[ht]
    		\centering
    		\begin{subfigure}{0.28\textwidth}
        		\centering
        		\includegraphics[width=\linewidth]{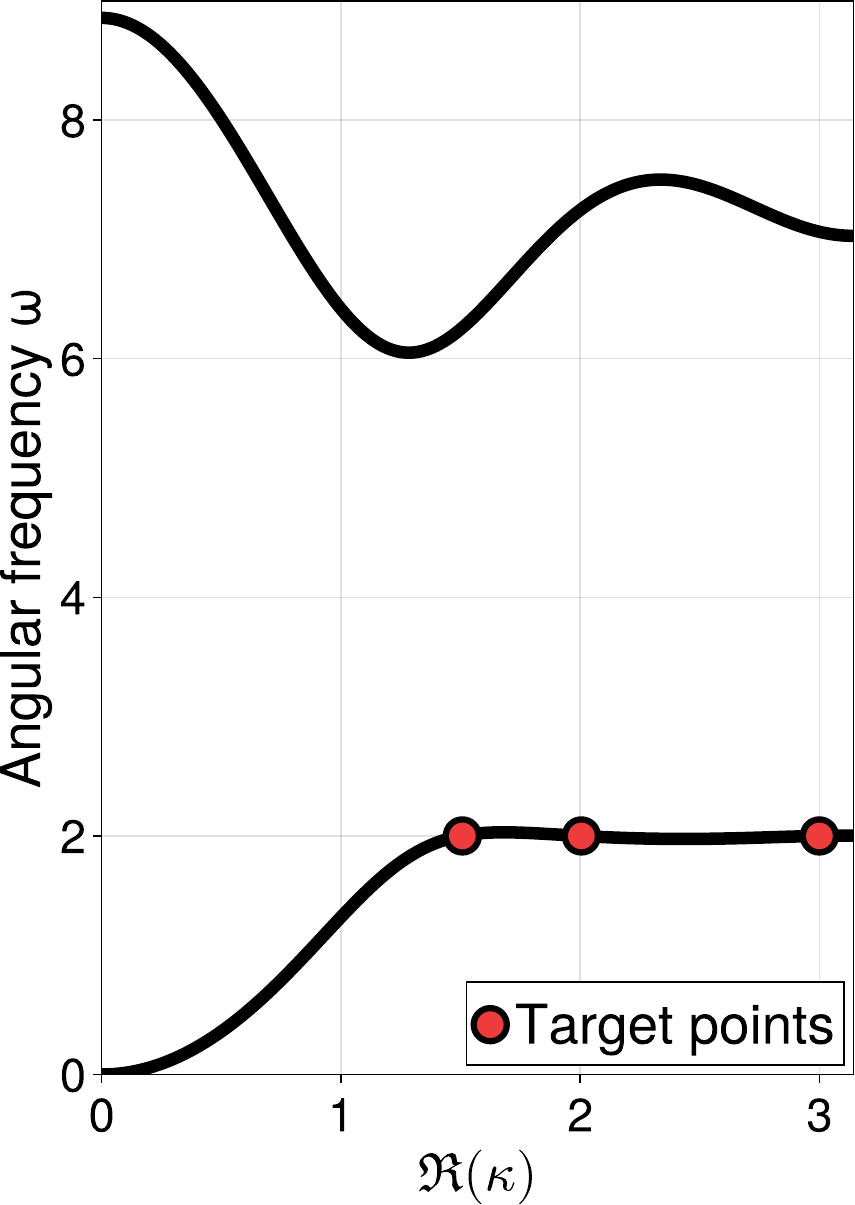}
        		\caption{}
				\label{fig:dispersion_beam_01}
    		\end{subfigure}
    		\hspace{0.02\textwidth} 
    		\begin{subfigure}{0.28\textwidth}
        		\centering
        		\includegraphics[width=\linewidth]{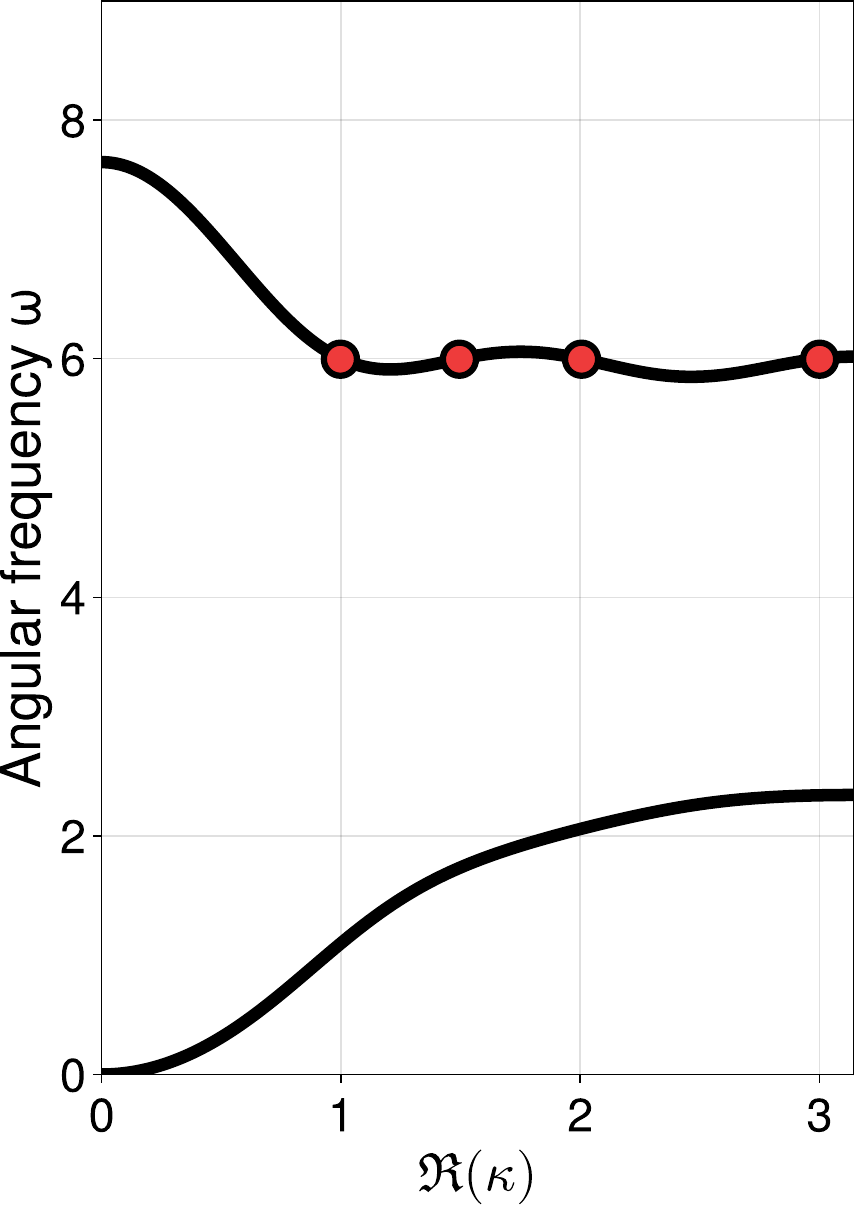}
        		\caption{}
    		\end{subfigure}
			\hspace{0.02\textwidth} 
			\begin{subfigure}{0.28\textwidth}
        		\centering
        		\includegraphics[width=\linewidth]{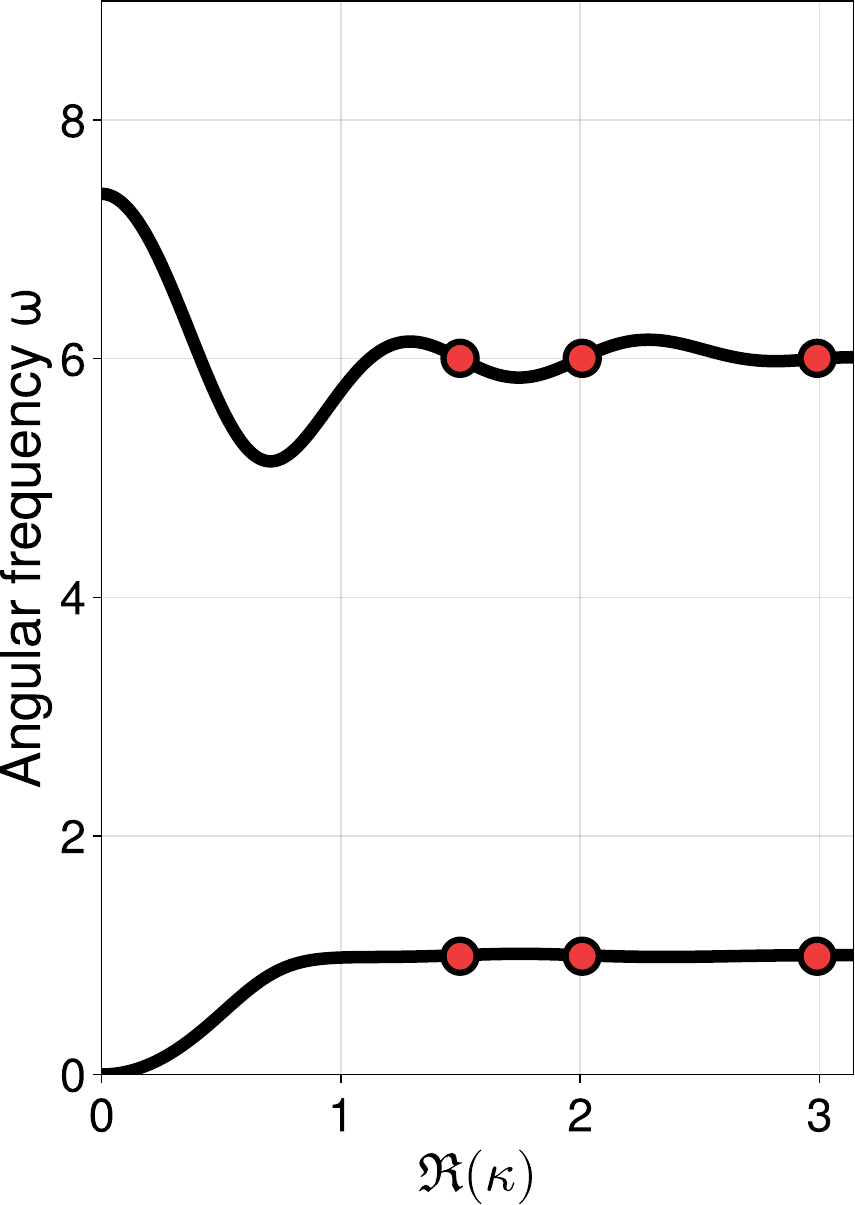}
        		\caption{}
    		\end{subfigure}
			\caption{Examples of dispersion curves for the Euler-Bernoulli beam lattice.  
			In the first two cases, a single branch is interpolated, while in the third case both branches are simultaneously constrained (see Table~\ref{table:dispersion_beam}).}
			\label{fig:dispersion_beam}
		\end{figure}

		\added[id = RB]{As an illustrative example, whose result is shown in Figure~\ref{fig:dispersion_beam_01}, 
		the objective is to obtain a flat acoustic branch over the interval $[1.5,\pi]$.
		In this case, the interpolation points are selected solely based on the desired dispersion characteristics,
        namely enforcing $\omega_{-} \approx 2$ over the prescribed wavenumber range.
		This is achieved by selecting three target points:}
		\[
			(\kappa_1=1.5, \omega_1=2.0), \quad
			(\kappa_2=2.0, \omega_2=2.0), \quad
			(\kappa_3=2.5, \omega_3=2.0).
		\]
		\added[id = RB]{leading to the nonlinear system}
		\begin{equation}
			\begin{cases}
				\omega_-(\kappa_1 = 1.5, \boldsymbol{\beta})^2 - 2.0^2 &= 0 	\\
				\omega_-(\kappa_2 = 2.0, \boldsymbol{\beta})^2 - 2.0^2 &= 0		\\
				\omega_-(\kappa_3 = 2.5, \boldsymbol{\beta})^2 - 2.0^2 &= 0
			\end{cases}
		\end{equation}
		\added[id = RB]{Choosing a nonlocal interaction order $P=3$ yields a square system, which in this case admits a solution involving only positive stiffness parameters.
		In the absence of a solution, a natural strategy consists in relaxing the system by considering an underdetermined formulation,
		obtained by selecting $P>3$. Examples of such cases are presented in Appendix~\ref{sec:other_dispersion_curves}.
		More generally, increasing the number of nonlocal interactions introduces additional design parameters, thereby enlarging the feasible solution set.}

		Once the system is formulated, it can be solved using numerical optimization methods, as detailed in Appendix~\ref{sec:positivity_constraint}.
		Within this framework, additional constraints, such as the positivity of each parameter $\beta_p$, can be readily incorporated.
		Two additional examples of such parameterizations are shown in Figure~\ref{fig:dispersion_beam}.

		In summary, the Euler-Bernoulli beam lattice provides a representative setting in which the interpolation framework can be explicitly implemented and tested.  
		The model highlights both the potential and the limitations of the method, while offering practical criteria for feasible dispersion shaping.  
		Building on this foundation, the next section explores how parametric interpolation can be employed to engineer a variety of distinctive wave phenomena.

	\section{Engineering wave phenomena via parametric interpolation}
	\label{part:Engineering_wave_phenomena}
	The previous sections established the theoretical framework for uniform nonlocal lattices and introduced a systematic interpolation method for tailoring their dispersion relations.  
	The Euler–Bernoulli beam lattice was then adopted as a representative model, providing explicit expressions for the interaction matrices and serving as a reference setting for practical implementation.  

	This section illustrates the potential of the proposed approach through a series of application examples.  
	By prescribing selected points on the dispersion curves, distinct and physically relevant wave phenomena can be engineered,
	ranging from qualitative modifications of branch topology to quantitative control of wave-packet dynamics.  
	Three representative cases are considered: the construction of rotons, the tuning of group velocity dispersion, and the design of evanescent modes within band gaps.  
	Together, these examples demonstrate the versatility of parametric interpolation as a tool for engineering wave propagation in nonlocal lattices.
		\subsection{Constructing rotons via parametric dispersion control}
		\label{part:Rotons}

		In the Euler–Bernoulli beam lattice restricted to local interactions, both branches of the dispersion relation are injective over $\kappa \in [0,\pi]$,  
		as can be observed from the analytical expressions derived in Appendix~\ref{sec:AnalyticalDispersionBeam}.
		For a given angular frequency $\omega$, the Floquet-Bloch condition~\eqref{eq:Floquet_form} admits a single propagation
		constant \added[id = Oth]{$| \lambda | \in (0,1]$} in the purely local case.  
		Introducing nonlocal interactions fundamentally changes this behavior: the dispersion branches may lose their injectivity,  
		allowing multiple propagation constants $\lambda$ to correspond to the same frequency $\omega$.
		\added[id = RB]{This manifests itself through the emergence of local extrema in the dispersion curves.
		Such dispersion profiles are commonly described as \emph{roton-like} in the literature~\citep{iglesias2021experimental,wang2022nonlocal,zhu2022observation,iorio2024roton}.}

\begin{figure*}[ht]
    		\centering
    		\begin{subfigure}{0.29\textwidth}
        		\centering
        		\includegraphics[width=\linewidth]{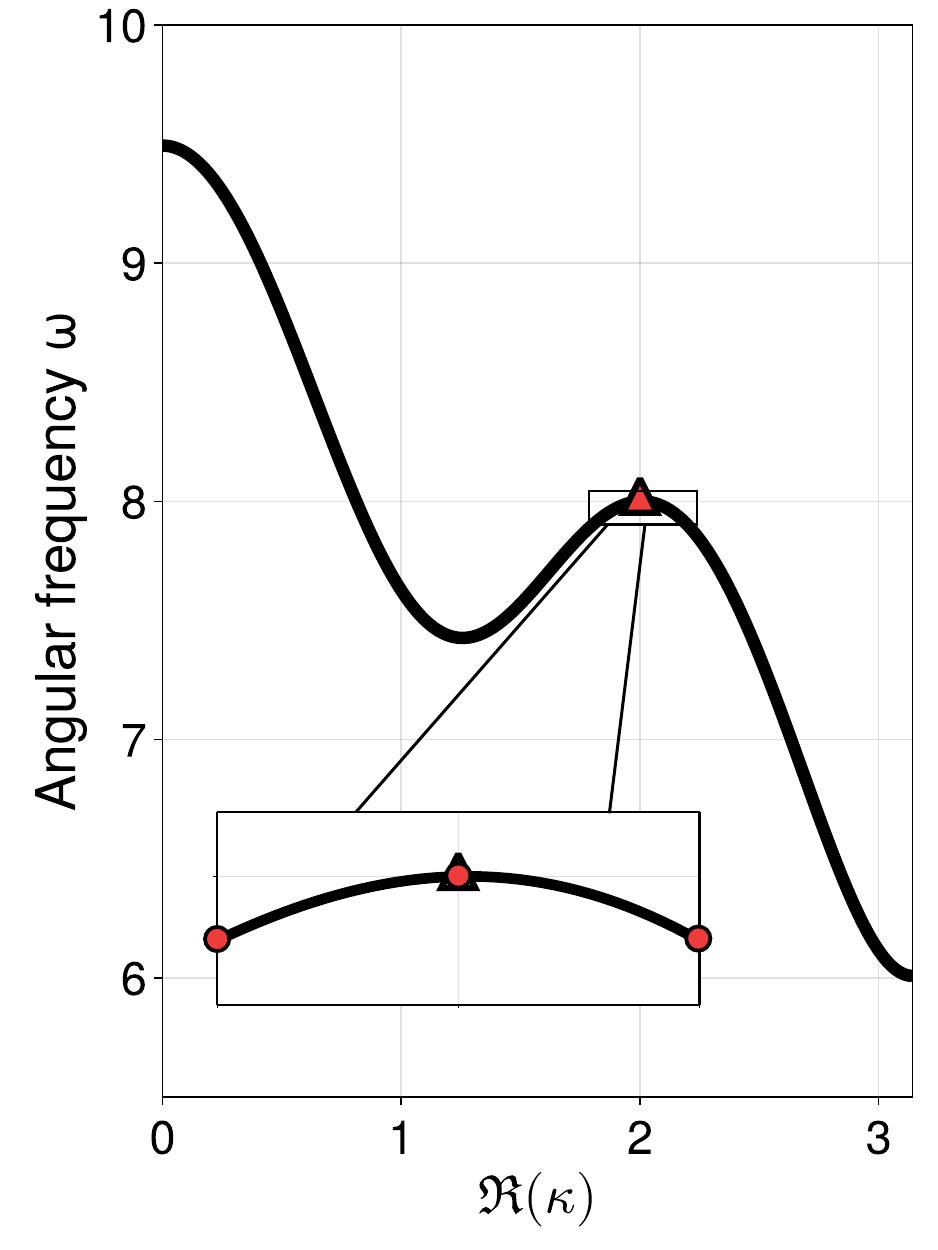}
        		\caption{}
				\label{fig:roton_disp02}
    		\end{subfigure}%
			\quad
    		\begin{subfigure}{0.29\textwidth}
        		\centering
        		\includegraphics[width=\linewidth]{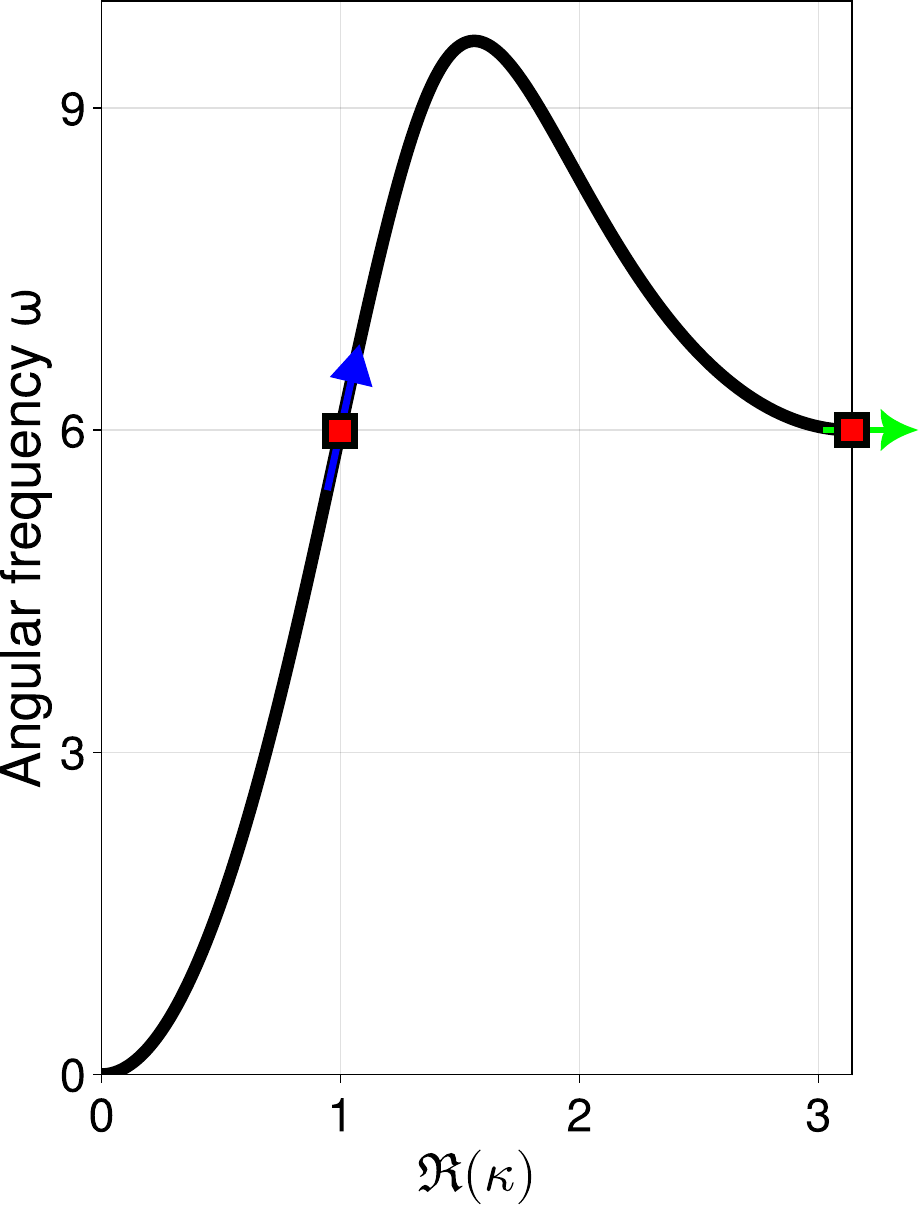}
        		\caption{}
				\label{fig:roton_disp01}
    		\end{subfigure}
    		\begin{subfigure}{0.29\textwidth}
        		\centering
        		\includegraphics[width=\linewidth]{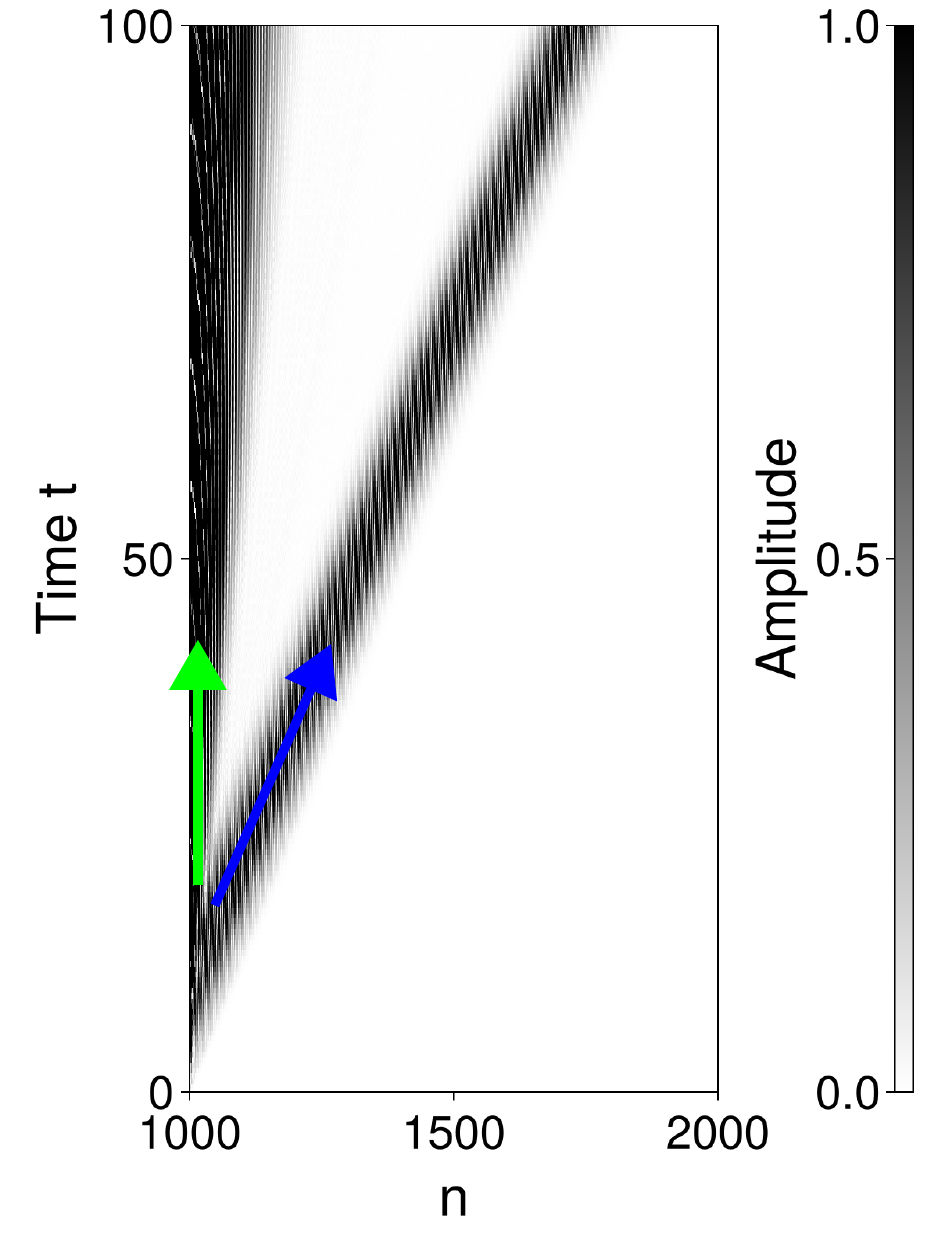}
        		\caption{}
				\label{fig:roton_transt}
    		\end{subfigure}

    		\caption{(\subref{fig:roton_disp02}) Example of a roton created at the frequency $\omega_{+}=8$. 
    		Three interpolation points are selected to enforce a vanishing central finite difference at $\omega_{+}$ (see Table~\ref{table:Rotons}). 
    		(\subref{fig:roton_disp01}) Model exhibiting two propagating solutions for the same frequency $\omega_{-}=6$,
			obtained by prescribing two interpolation points at $(\kappa_{1}=1, \omega_{-})$ and $(\kappa_{2}=\pi, \omega_{-})$. 
    		(\subref{fig:roton_transt}) Transient response associated with (\subref{fig:roton_disp01}):
			the excitation at $\omega_{-}=6$ splits into two wave packets with distinct group velocities.}
			\label{fig:rotons_dispersion}
		\end{figure*}

		The parametrization method enables the systematic construction of such non-injective dispersion curves.
		\added[id = RB]{Two examples are considered. First, a parametrization is introduced to enforce a vanishing group velocity at a prescribed frequency $\omega_{0}$.
		In particular, the group velocity associated with a given frequency can be controlled through the specification of three interpolation points.}
		Three interpolation points $(\lambda_{-1},\omega_{-1})$, $(\lambda_{0},\omega_{0})$, and $(\lambda_{+1},\omega_{+1})$ are chosen symmetrically around $\kappa_0$,
		with spacing $\Delta\kappa$:  
		\[
			\omega_{-1} < \omega_{0} < \omega_{+1}, \quad 
			\kappa_{-1} < \kappa_0 < \kappa_{+1}, \quad 
			\kappa_{0} - \kappa_{-1} = \kappa_{+1} - \kappa_{0} = \Delta \kappa .
		\]
		A vanishing slope at $(\lambda_0,\omega_0)$ is then imposed through the central finite-difference criterion,  
		\begin{equation}
			\frac{\omega_{+1} - \omega_{-1}}{2\Delta \kappa} = 0.
		\end{equation}
		To ensure that the point indeed corresponds to a roton, the curvature must be nonzero, which is verified through the second-order finite difference:   
		\begin{equation}
			\omega_{-1} - 2\omega_0 + \omega_{+1} \neq 0,
		\end{equation}  
		positive for a local minimum and negative for a local maximum.
		The dispersion curve of a model parametrized to exhibit a local minimum at $(\kappa = 2.0, \omega_+ = 8)$ is shown in Figure~\ref{fig:roton_disp02}.

		\added[id = RB]{A second possibility for designing a non-injective dispersion curve is to associate multiple distinct propagation constants
		with a prescribed frequency $\omega_0$. Within the interpolation framework, this amounts to selecting
		several target points $(\lambda_i, \omega_0)$ with distinct propagation constants.} 
		The resulting system~\eqref{eq:system_interpolation} then yields a lattice model whose dispersion relation admits multiple propagation constants at the same angular frequency.  
		The dispersion curve of such a system is presented in Figure~\ref{fig:roton_disp01}.
		In this example the condition $\partial_\kappa \omega(\kappa_1) \neq \partial_\kappa \omega(\kappa_2)$,
		corresponds to distinct group velocities associated with $\omega_0$.
		The implications of this property can be highlighted through transient simulations.  
		Consider a Gaussian wave packet imposed on the reference node $x_0$, centered around the frequency $\omega_0$:  
		\begin{equation}
			u_0(t) = A \cos\!\big(\omega_0 (t - t_0)\big) \, e^{-((t-t_0)/\tau)^2}.
		\label{eq:wave_packet}
		\end{equation}
		As shown in Figure~\ref{fig:roton_transt},
		the initially localized excitation evolves into two distinct packets that propagate with the two group velocities determined at $\omega_0$.

		The emergence of rotons illustrates how parametric interpolation can modify the qualitative structure of dispersion relations.
		The same principle can be extended to tune more subtle features, such as the group velocity dispersion, which governs the spreading of wave packets over time.

	\subsection{Tailoring group velocity dispersion via parametric interpolation}
		In discrete Euler-Bernoulli beam lattices, the relation between angular frequency $\omega$ and wavenumber $\kappa$ is inherently nonlinear,
		reflecting their dispersive nature.  
		As a consequence, a wave packet centered around a given frequency $\omega_0$ gradually spreads over time due to dispersion effects.  
		This spreading is governed by the \emph{group velocity dispersion} (GVD), defined as
		\begin{equation}
			\mathbf{GVD}(\omega_0) = \partial_\omega^2 \kappa(\omega_0)
		\end{equation}
		By applying the local inversion theorem, the GVD can be expressed in terms of the curvature of the dispersion relation:
		\begin{equation}
			\mathbf{GVD}(\omega_0)  = - \frac{\partial_{\kappa}^2 \omega(\kappa_0)}{c_0^2}
		\end{equation}
		where $c_0 = \partial_\kappa \omega(\kappa_0)$ denotes the group velocity at $\kappa_0$.  
		This expression highlights the central role played by the second derivative of $\omega(\kappa)$ in controlling the spreading of wave packets.

\begin{figure}[ht]
    		\centering
    		\begin{subfigure}{0.28\textwidth}
        		\centering
        		\includegraphics[width=\linewidth]{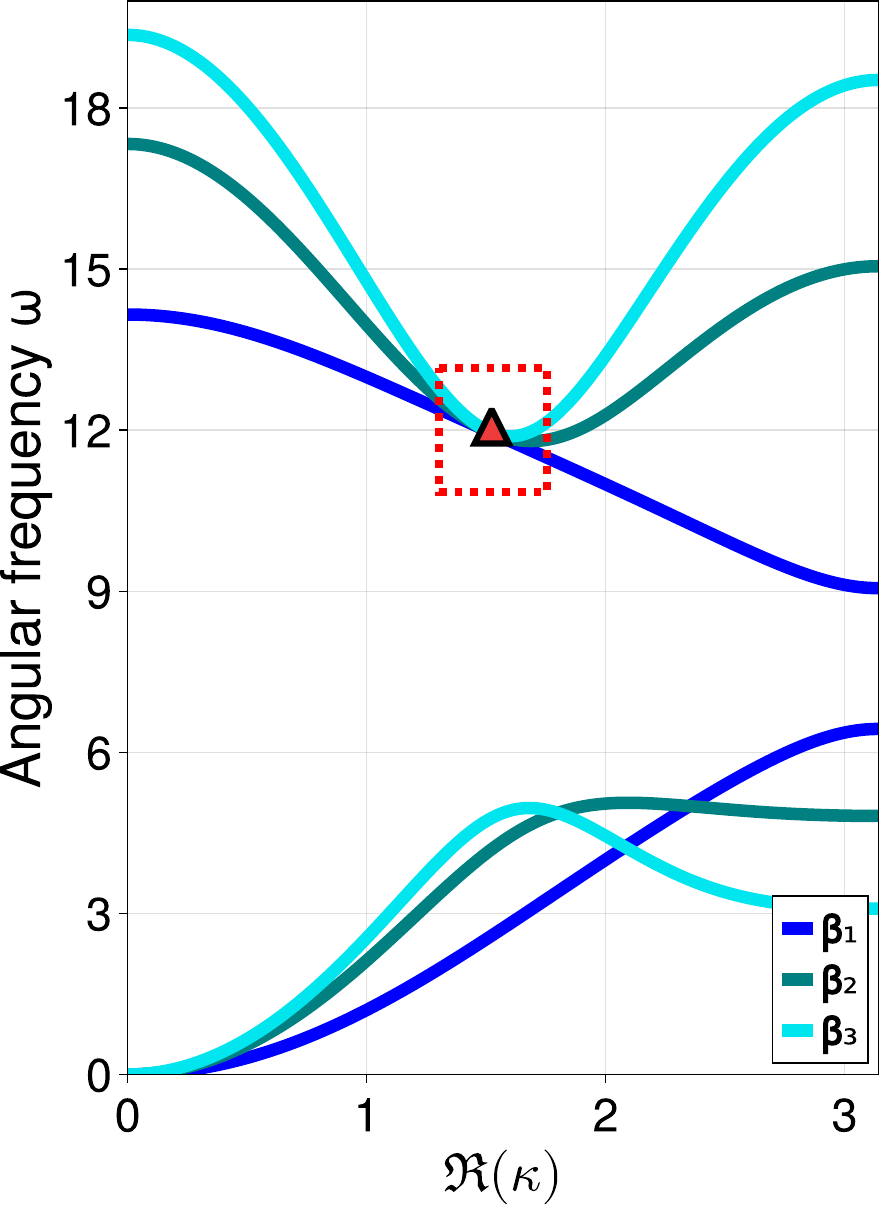}
        		\caption{}
				\label{fig:GVD_01}
    		\end{subfigure}
    		\hspace{0.02\textwidth} 
    		\begin{subfigure}{0.39\textwidth}
        		\centering
        		\includegraphics[width=\linewidth]{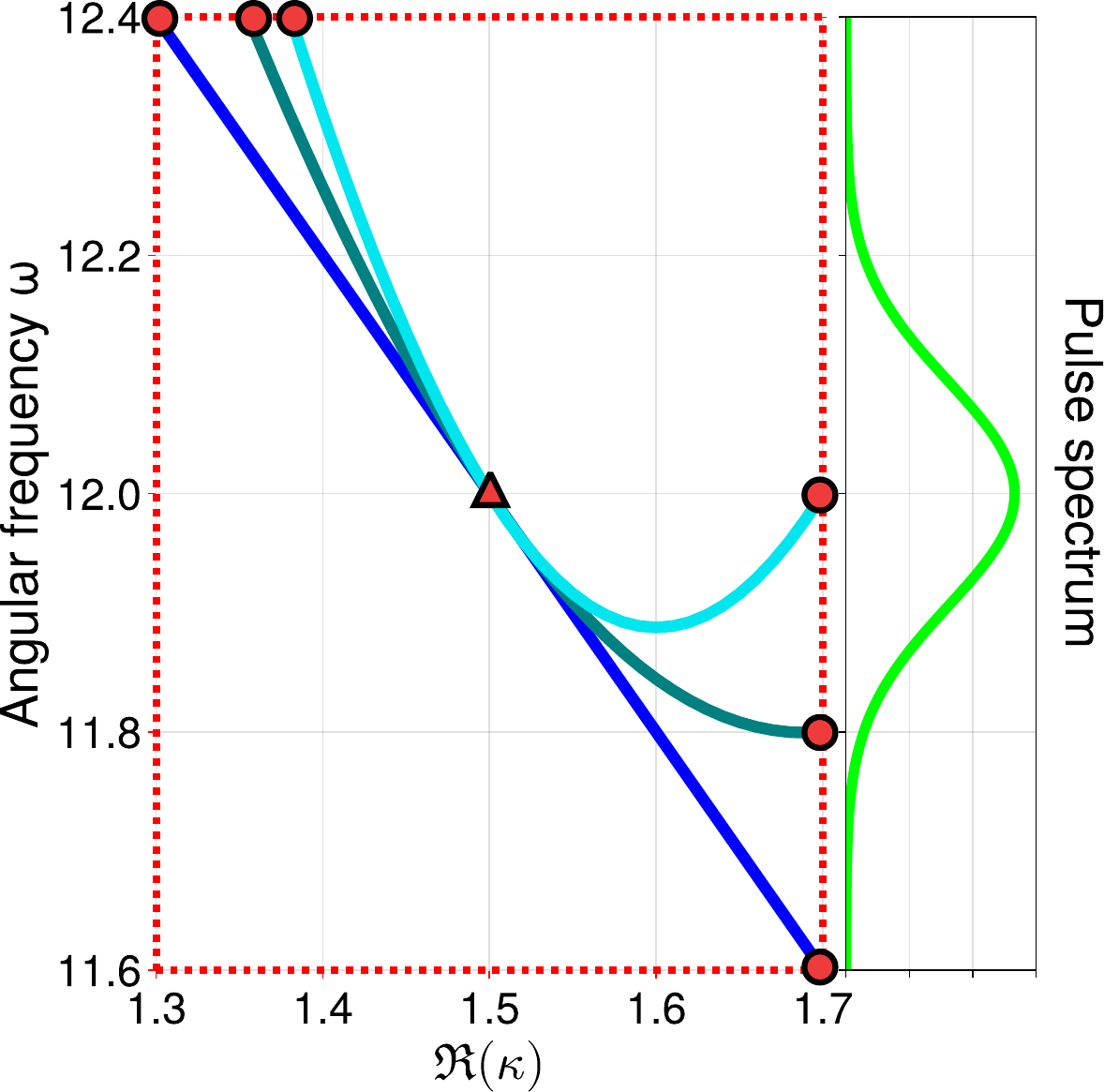}
        		\caption{}
				\label{fig:GVD_02}
    		\end{subfigure}
			\hspace{0.02\textwidth} 
			\begin{subfigure}{0.208\textwidth}
        		\centering
        		\includegraphics[width=\linewidth]{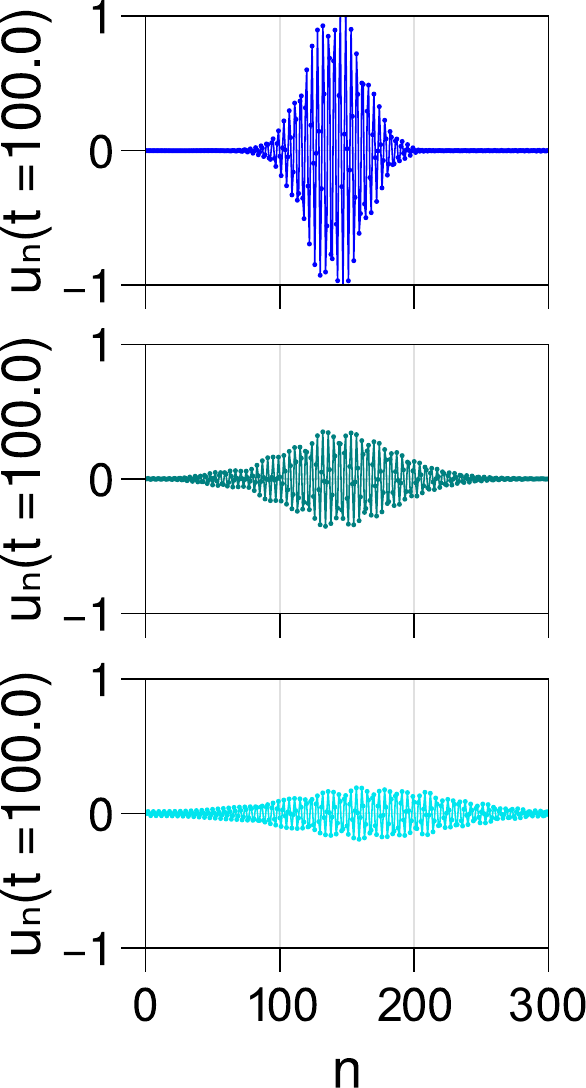}
        		\caption{}
				\label{fig:GVD_03}
    		\end{subfigure}
    		\caption{(\subref{fig:GVD_01}) Dispersion curves of three parametrized models, each tailored in a frequency range around $\omega_0 = 12.0$.  
			(\subref{fig:GVD_02}) Zoom on the customized region.
			Three interpolation points are selected to simultaneously control the slope and curvature of the dispersion relation near $\omega_0 = 12.0$ (see Table~\ref{table:GVD}).  
			For the three models, the group velocity is fixed at $c_0 = -2$, while the group velocity dispersion (GVD) takes values $0$, $1.25$, and $2.5$, respectively.  
			(\subref{fig:GVD_03}) Transient wave-packet analysis.  
			The excitation imposed on node $x_0$ follows Eq.~\eqref{eq:wave_packet}.  
			The corresponding spectral content, shown in green in~(\subref{fig:GVD_02}), is centered around $\omega_0$ and lies within the parametrized region of the dispersion curves.}
    		\label{fig:GVD_curve}
		\end{figure}

		As illustrated in Section~\ref{part:Rotons},
		the interpolation method can be employed to prescribe not only the slope of the dispersion relation at selected points but also its local curvature.  
		By selecting three interpolation points in the neighborhood of $(\kappa_0, \omega_0)$ and applying a finite-difference criterion,
		both the group velocity and the GVD can be tuned independently.  
		It therefore becomes possible, for a given angular frequency $\omega_0$, to design lattices in which wave packets experience stronger or weaker spatial spreading over time.  
		An example of such a parameterization is shown in Figure~\ref{fig:GVD_curve}:
		after locally tuning the slope and curvature of the dispersion curves of three different models at the same frequency $\omega_0$,  
		the corresponding wave packets centered around $\omega_0$ are observed to spread in space at different rates, depending on the imposed curvature.

		These results show that parametric interpolation not only alters the qualitative topology of dispersion branches,
		as in the case of rotons, but also provides a systematic means of adjusting quantitative properties such as the group velocity dispersion.
		This flexibility in shaping propagating-wave dynamics naturally extends to evanescent modes within band gaps,
		which can likewise be parametrically controlled, as discussed in the next subsection.

		\subsection{Designing evanescent wave decay via parametric interpolation}
		For a fixed angular frequency $\omega_0$, the interpolation method introduced earlier can be used to identify
		wave solutions associated with propagation constants $\lambda$ of unit modulus.  
		Such solutions correspond to propagating waves that maintain constant amplitude in the absence of dissipative effects.  
		However, even without material dissipation, the periodicity of the lattice gives rise to frequency intervals
		in which admissible propagation constants satisfy $|\lambda| \neq 1$.
		The corresponding modes, referred to as \emph{evanescent waves}, exhibit exponential spatial decay along a preferred direction.
	
		The propagation constant can be expressed in polar form as
		\begin{equation}
    		\lambda = |\lambda| e^{-i l\kappa_R},
		\end{equation}
		or equivalently,
		\begin{equation}
    		\lambda = e^{-i l\kappa}, 
		\end{equation}
		where $\kappa = \kappa_R + i \kappa_I$ and $\kappa_I = \ln(|\lambda|)$.  
		When $|\lambda| < 1$, the amplitude decays to the right with factor $e^{n \kappa_I}$.  
		Moreover, if $\lambda$ is a solution at frequency $\omega_0$, then $1/\lambda$ is also a solution, so the analysis may,
		without loss of generality, be restricted to $|\lambda| < 1$.

\begin{figure*}[ht]
			\centering
			\includegraphics[scale=0.32]{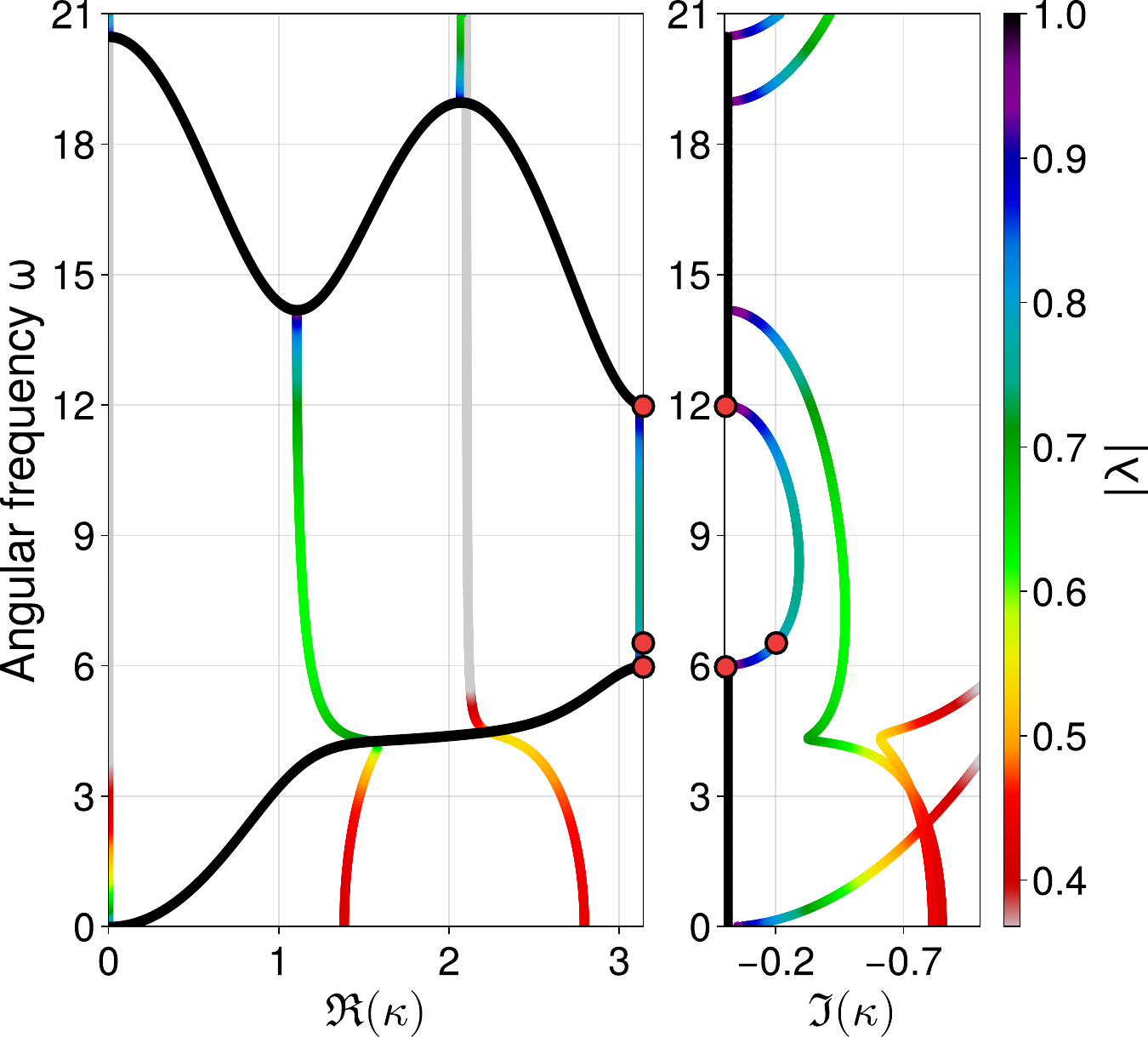}
			\includegraphics[scale=0.32]{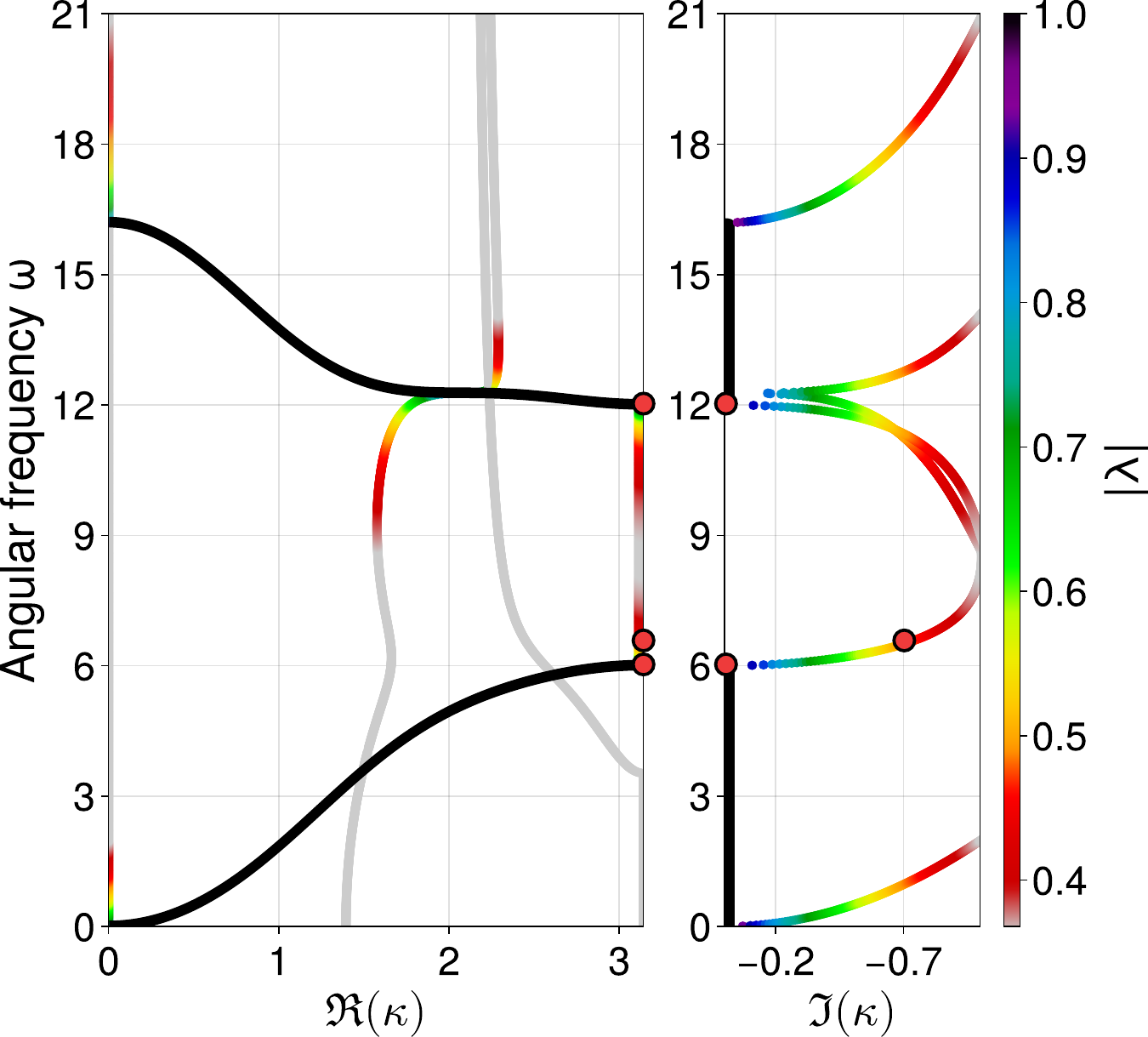}
			\vskip\baselineskip
			\includegraphics[scale=0.28]{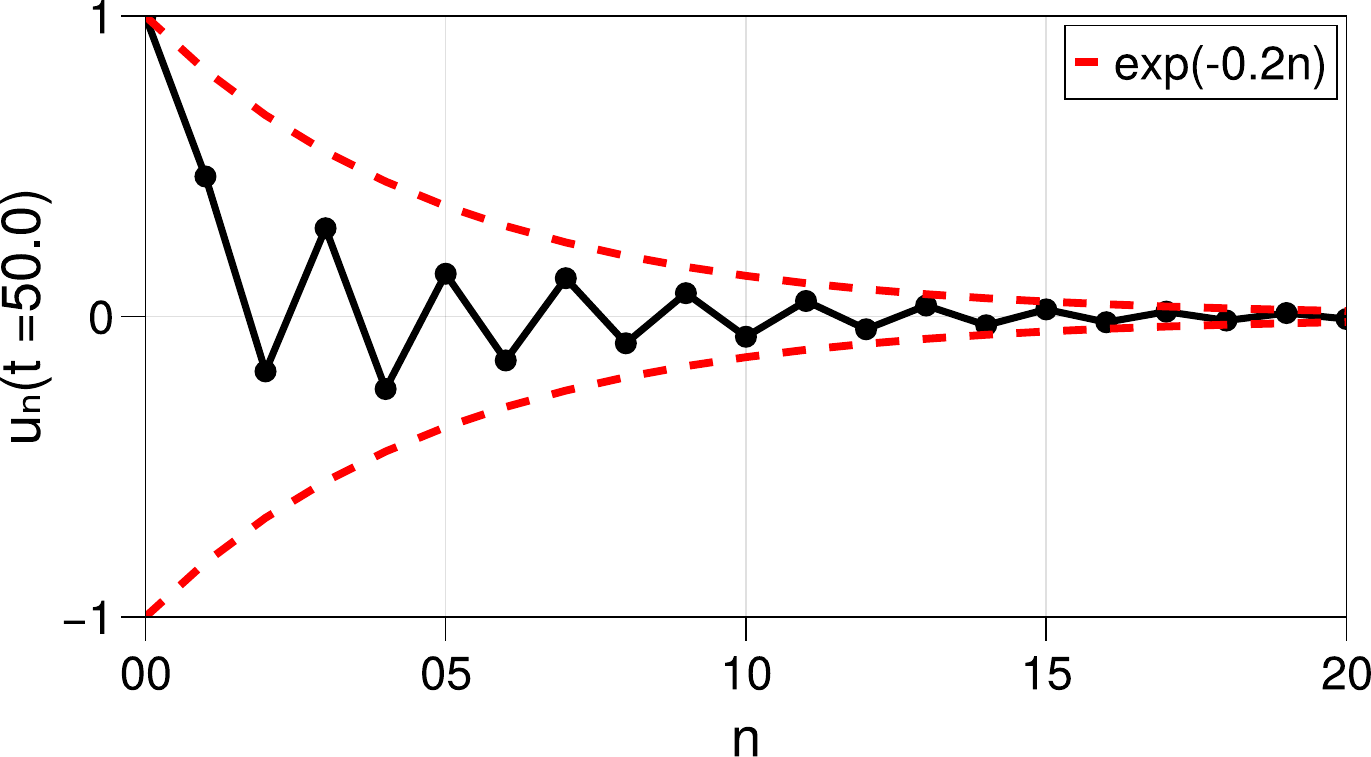}
			\hspace{0.04\textwidth} 
			\includegraphics[scale=0.28]{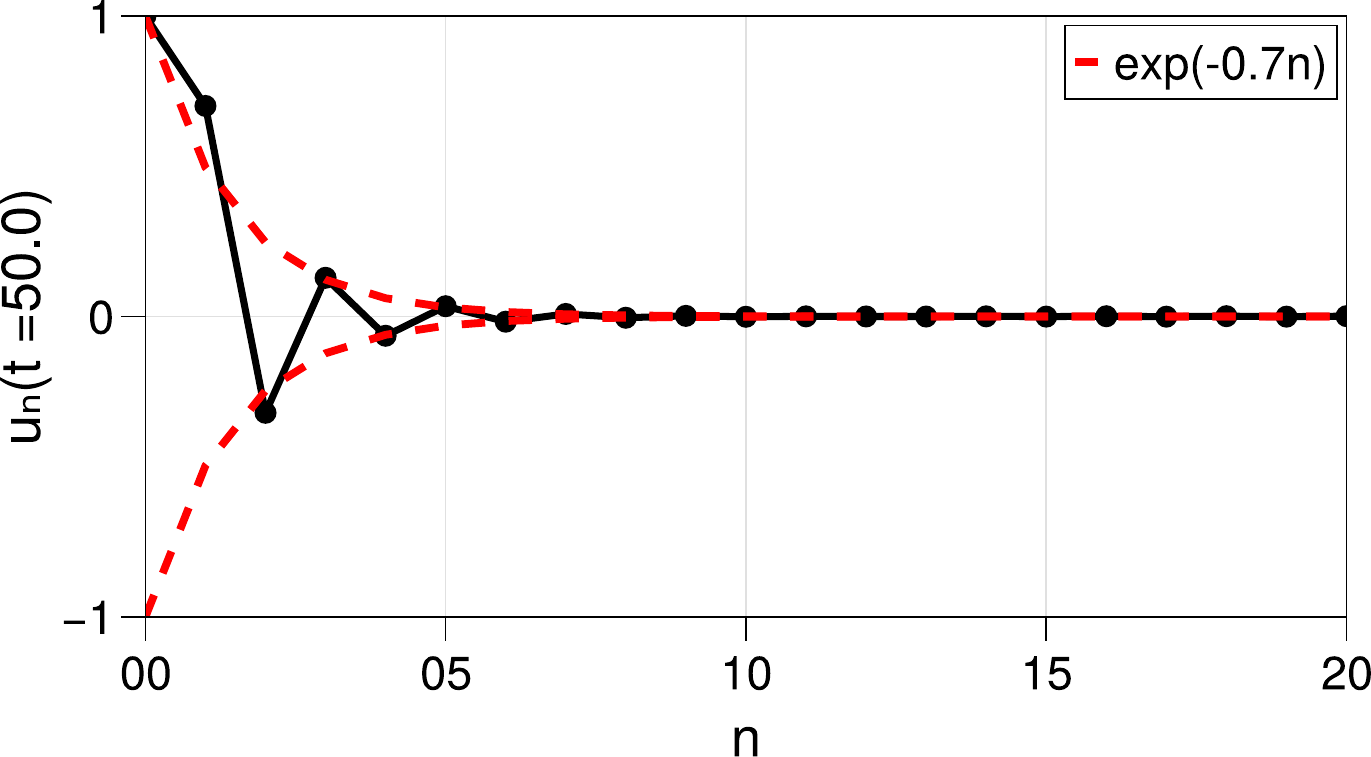}
		\caption{Two uniform nonlocal beam-lattice models exhibiting the same band gap $[6.0, 12.0]$.  
		The colormap represents the modulus of $\lambda$, providing a complementary visualization of $\Im(\kappa)$.  
		An evanescent wave solution at $\omega = 6.5$ is parametrized with different decay rates: $\kappa_I = -0.2$ in the first case and $\kappa_I = -0.7$ in the second.  
		Transient analysis confirms that the spatial attenuation follows approximately $e^{\kappa_I}$.  
		The excitation applied to the reference node $x_0$ follows Eq.~\eqref{eq:wave_packet}.  
		The parameters used are listed in Table~\ref{table:evanescent_wave}.}
		\label{fig:evanescent_dispersion}
		\end{figure*}

		The proposed interpolation framework naturally extends to such evanescent solutions by solving the dispersion relation in the complex plane. 
		For the Euler-Bernoulli beam lattice, the dispersion relation exhibits two propagating branches separated by a band gap.  
		Within this interval, the only admissible solutions correspond to evanescent modes.  
		The band-gap boundaries $[\omega_0, \omega_1]$ can be determined by selecting two target points, $(\lambda_0 = -1, \omega_0)$ and $(\lambda_1 = -1, \omega_1)$.  
		Figure~\ref{fig:evanescent_dispersion} shows the dispersion curves of two models parametrized
		using these two target points so as to produce the same band gap over $[\omega_1, \omega_2]$.  
		A third target point, which differs between the two models, $(\lambda_3, \omega_3)$ with $|\lambda_3| < 1$ and $\omega_3 \in [\omega_1, \omega_2]$,  
		is introduced to construct a parametrized evanescent solution within the band gap.  
		The modulus of $\lambda_3$ directly controls the rate of exponential decay of the corresponding mode.

		This example shows that evanescent modes can be actively designed and tailored within the proposed framework.  
		Such control over spatial decay complements the previous results on rotons and group velocity dispersion,
		thereby further expanding the range of achievable wave-propagation behaviors in nonlocal lattices.

	\clearpage
	\section{Conclusion}
	This work introduced a systematic interpolation-based method for tailoring dispersion relations in uniform nonlocal lattices.  
	By prescribing selected frequency-wavenumber points as interpolation constraints,
	the method provides localized and flexible customization without requiring closed-form dispersion expressions or global curve-fitting procedures.  
	The approach was developed in a general theoretical setting and illustrated through the Euler-Bernoulli beam lattice,
	a representative model capturing both bending and rotational dynamics.

	Several applications were presented to highlight the range of achievable phenomena.  
	The construction of roton-like extrema demonstrated how qualitative topological features of dispersion branches can be engineered.  
	The adjustment of group velocity dispersion showed that quantitative wave-packet spreading can be controlled through local curvature tuning.  
	Finally, the parametrization of evanescent modes within band gaps illustrated that exponential decay rates can likewise be designed,
	extending the scope of dispersion control beyond propagating waves.

	Together, these examples establish parametric interpolation as a versatile and computationally tractable tool for dispersion engineering in nonlocal lattices.  
	Future work may extend the framework to higher-dimensional systems, explore the inclusion of damping and dissipation mechanisms,
	or integrate the method with optimization schemes for application-driven design.  
	Such developments would further enhance the applicability of nonlocal lattices in advanced metamaterial concepts for vibration isolation,
	acoustic control, and wave-based technologies.

	\section*{Acknowledgements}
	The authors acknowledge the French National Research Agency (ANR) for its financial support through the JCJC-Archi-Noise project (grant no.~ANR-23-CE51-0045).

	\newpage
	\appendix
	\section{Appendix}

		\subsection{General formulation of the dispersion relation in uniform nonlocal models}
		\label{part:Dispersion_Relation}
		The derivation of the dispersion relation for uniform nonlocal lattices, presented here, provides the theoretical foundation for the tuning procedure 
		introduced in Section~\ref{sec:tuning}.  
		It formalises the connection between the physical parameters $\beta$ and the spectral properties $(\lambda, \omega)$ of the system, 
		thus enabling the targeted adjustment of dispersion curves.
		
		Consider a uniform nonlocal lattice of order $P$.  
		Such a system is described by specifying the interactions between its constituent nodes,  
		where each interaction depends solely on its nonlocal order $p \in \{1, \dots, P\}$.  
		
		Let $x_l$ and $x_r$ denote two nodes separated by a distance $pl$.  
		The equations of motion for this pair can be expressed in block-matrix form as
		\begin{equation}
			\begin{bmatrix}
				\mathbf{M}_p^+	& \mathbf{O}_p^+	\\
				\mathbf{O}_p^-	& \mathbf{M}_p^-
			\end{bmatrix}
			\partial_t^2
			\begin{bmatrix}
				\mathbf{u}_l		\\
				\mathbf{u}_r
			\end{bmatrix}
			+
			\begin{bmatrix}
				\mathbf{A}_p^+		& \mathbf{B}_p^+	\\
				\mathbf{B}_p^- 	& \mathbf{A}_p^-
			\end{bmatrix}
			\begin{bmatrix}
				\mathbf{u}_l		\\
				\mathbf{u}_r
			\end{bmatrix}
			=
			\mathbf{0}
		\label{eq:motion_interaction}
		\end{equation}
		where the matrices  $\mathbf{M}_p^\pm$, $\mathbf{O}_p^\pm$, $\mathbf{A}_p^\pm$, and $\mathbf{B}_p^\pm$  
		are $d \times d$ blocks determined by the physical parameter $\beta_p$.
		By assembling two such elements, the linear dependence of a given node $x_n$ on its $p$-th order neighbors $x_{n-p}$ and $x_{n+p}$ can be expressed as
		\begin{equation}
			\begin{bmatrix}
				\mathbf{O}_p^- 		&\mathbf{M}_p^- +\mathbf{M}_p^+		&\mathbf{O}_p^+	
			\end{bmatrix}
			\partial_t^2
			\begin{bmatrix}
				\mathbf{u}_{n-p}			\\
				\mathbf{u}_n				\\
				\mathbf{u}_{n+p}
			\end{bmatrix}
			+
			\begin{bmatrix}
				\mathbf{B}_p^- 		&\mathbf{A}_p^- +\mathbf{A}_p^+		&\mathbf{B}_p^+	
			\end{bmatrix}
			\begin{bmatrix}
				\mathbf{u}_{n-p}			\\
				\mathbf{u}_n				\\
				\mathbf{u}_{n+p}
			\end{bmatrix}
			=
			\mathbf{0}
		\end{equation}
		
		Continuing this assembly process for all $p \leq P$ yields the stiffness matrix $\mathbf{K}$ from Eq.~\eqref{eq:motion_nonlocal_model} in the compact form			 
		\begin{equation}
			\mathbf{K} = 
			\begin{bmatrix}
				\mathbf{B}_{P}^-		& \mathbf{B}_{P-1}^-		& \dots		& \mathbf{A} 	& \dots		& \mathbf{B}_{P-1}^+		& \mathbf{B}_{P}^+
			\end{bmatrix}
		\end{equation}
		where
		\begin{equation}
			\mathbf{A} = \sum_{p=1}^P (\mathbf{A}_p^- +\mathbf{A}_p^+)
		\end{equation}
		A similar construction applies to the mass matrix $\mathbf{M}$.
		In the present work, a lumped-mass assumption is adopted, namely $\mathbf{O}_p^\pm = \mathbf{0}$ and $\mathbf{M}_p^- = \mathbf{M}_p^+$ are diagonal.
		
		Seeking a Floquet-Bloch wave solution of Eq.~\eqref{eq:motion_nonlocal_model} in the form of Eq.~\eqref{eq:Floquet_form} gives
		\begin{equation}
			-\omega^2 \mathbf{M}\,\mathbf{\Phi}\,u_n + \mathbf{K}\,
			\begin{bmatrix}
				\lambda^{-P}\mathbf{\Phi} &	\dots & \mathbf{\Phi} &	\dots 	& \lambda^{-P}\mathbf{\Phi}
			\end{bmatrix}^T
			u_n = 0 
		\end{equation}
		This directly leads to the dispersion relation given in Eq.~\eqref{eq:nonlocal_unif_dispersion}.
	
		\subsection{Monoatomic chain with uniform nonlocal interactions}
		\label{part:monoatomic-lattice}
		
		This subsection considers a particular case of the uniform nonlocal lattice model.
		The system consists of a one-dimensional periodic chain of identical nodes, each having a single degree of freedom and mass $m$,
		separated by a constant lattice spacing $l$.
		
		The mechanical interactions between nodes are purely axial and described by springs.  
		Nearest-neighbor couplings ($p=1$) have stiffness $\beta_1$,  
		while higher-order nonlocal interactions connect each node to its $p$-th neighbors ($p = 2, \dots, P$) on both sides, with stiffness $\beta_p$. 
		
		Under these assumptions, the general dispersion relation  
		\begin{equation}
			m\,\omega^2 = 2\sum_{p=1}^P \beta_p (1- \cos(p\kappa))
		\label{eq:dispersion_monoatomic}
		\end{equation}
		Using the trigonometric identity  $1 - \cos(p\kappa) = 2 \sin^2 \left( p\kappa/2 \right)$,  Eq.~\eqref{eq:dispersion_monoatomic} becomes
		\begin{equation}
			m \, \omega^2(\kappa) = 4 \sum_{p=1}^P \beta_p \, \sin^2\left( p\kappa/2 \right).
		\label{eq:dispersion_monoatomic_sin}
		\end{equation}

		Following the tuning approach presented in Section~\ref{sec:tuning},  a set of $P$ wave numbers is selected:
		\begin{equation}
    			\mathbf{k} = [\kappa_1, \dots, \kappa_P],
		\end{equation}
		together with the corresponding target squared angular frequencies:
		\begin{equation}
			\boldsymbol{\omega}^2 = [\omega_1^2, \dots, \omega_P^2].
		\end{equation}
		Substituting these values into Eq.~\eqref{eq:dispersion_monoatomic_sin}  leads to a linear system in the unknown stiffnesses $\beta_p$:
		\begin{equation}
			\mathbf{A} \, \boldsymbol{\beta} = m \, \boldsymbol{\omega}^2,
		\label{eq:linear_system_monoatomic}
		\end{equation}
		where $\boldsymbol{\beta} = [\beta_1, \dots, \beta_P]^T$  and the matrix $\mathbf{A} \in \mathbb{R}^{P\times P}$ is defined by
		\begin{equation}
			\mathbf{A} = 4
			\begin{bmatrix}
				\sin^2(\kappa_1/2)	& \dots 		& \sin^2(P\kappa_1/2)		\\
				\vdots				& \ddots		& \vdots					\\
				\sin^2(\kappa_P/2)	& \dots 		& \sin^2(P\kappa_P/2)		\\
			\end{bmatrix}.
		\end{equation}
		The solution of Eq.~\eqref{eq:linear_system_monoatomic} provides the stiffness values $\beta_p$ ensuring that the dispersion curve 
		passes exactly through the prescribed $(\kappa_i, \omega_i)$ points. 
		To guarantee physical admissibility ($\beta_p \geqslant 0$),  the system can be solved using a non-negative least squares (NNLS) approach.

		\subsection{Euler-Bernoulli beam lattices}

		\subsubsection{Dispersion relation for the Euler-Bernoulli beam lattice}
		\label{sec:DispersionBeamLattices}
		
		This appendix derives the equations of motion for the Euler–Bernoulli beam lattice model introduced in Section \ref{section:Euler-BernoulliBeamLattices}.
		The starting point is the governing equation for a single Euler–Bernoulli beam, as given in \citep{graff2012wave}:
		\begin{equation}
			EI \frac{\partial^4 w}{\partial x^4} + \rho A \frac{\partial^2 w}{\partial t^2} = 0
		\end{equation}
		where $E$ denotes the Young's modulus, $I$ the second moment of inertia, $\rho$ the density, and $A$ the cross-sectional area.
		
		Consider an element of this beam of length $p$, discretized using standard finite element theory.  
		Each end of the element possesses two degrees of freedom: the transverse displacement $w$ and the cross-section rotation $\theta$.  
		Let $\mathbf{u}_l$ and $\mathbf{u}_r$ denote the nodal displacement--rotation vectors at the left and right ends, respectively.  
		The discrete equation of motion reads:
		\begin{equation}
			\mathbf{M}_e \,\partial_t^2 \,
			\begin{bmatrix}
				\mathbf{u}_l		\\
				\mathbf{u}_r
			\end{bmatrix}
			+ \mathbf{K}_e
			\begin{bmatrix}
				\mathbf{u}_l		\\
				\mathbf{u}_r
			\end{bmatrix}
			= \mathbf{0}
		\label{eq:element_poutre}
		\end{equation}
		with the stiffness matrix 
		\begin{equation}
			\mathbf{K}_e = \frac{EI}{p^3}
			\begin{bmatrix}
				12 		& 6p 		& -12 		& 6p \\
				6p 		& 4p^2 		& -6p 		& 2p^2 \\
				-12 		& -6p 		& 12 		& -6p \\
				6p 		& 2p^2 		& -6p 		& 4p^2
			\end{bmatrix}
		\end{equation}
		and the consistent mass matrix
		\begin{equation}
			\mathbf{M}_e = \frac{\rho A p}{420}
			\begin{bmatrix}
				156 		& 22p 		& 54 		& -13p \\
				22p 		& 4p^2 		& 13p 		& -3p^2 \\
				54 		& 13p 		& 156 		& -22p \\
				-13p 	& -3p^2 		& -22p 		& 4p^2
			\end{bmatrix}
		\end{equation}
		In the present work, masses are concentrated at the lattice interfaces.  
		Following the approach of \citet{madine2021dynamic}, a lumped mass matrix is adopted:
		\begin{equation}
			\mathbf{M}_e = \text{diag}
				\begin{bmatrix}
					1		& \mu 	\\
					1		& \mu
				\end{bmatrix}
		\end{equation}
		with $0<\mu\ll 1$ a small rotational-to-translational inertia ratio.
		\added[id = Oth]{Throughout this paper, the value $\mu = 0.06$ is adopted.}
		
		Equation~\eqref{eq:element_poutre} matches the general interaction form~\eqref{eq:motion_interaction}.  
		By direct identification, the interaction matrices for the Euler--Bernoulli beam lattice are obtained as those given in Section~\ref{section:Euler-BernoulliBeamLattices}.  
		This establishes the explicit link between the beam element formulation and the dispersion relation used in the main text.

		\subsubsection{Analytical dispersion relation for the Euler-Bernoulli beam lattice}
		\label{sec:AnalyticalDispersionBeam}

		In this appendix, an explicit analytical form of the dispersion relation for the Euler--Bernoulli beam lattice is derived.
		This formulation enables the interpolation system~\eqref{eq:system_interpolation} to be expressed in closed form for this specific case.
		The derivation starts from the definitions of the interaction matrices $\mathbf{A}_p^-$, $\mathbf{A}_p^+$, $\mathbf{B}_p^-$,
		and $\mathbf{B}_p^+$ given in Section~\ref{section:Euler-BernoulliBeamLattices}. The total stiffness contribution can be written as
		\begin{equation}
		 	\sum_{p=1}^P \mathcal{K}_p(\lambda, \beta_p) =
		 	\sum_{p=1}^P \frac{\beta_p}{p^3}
		 	\left(
		 	\begin{bmatrix}
		 		24									& 0								\\
		 		0									& 8
		 	\end{bmatrix}
		 	+
		 	\begin{bmatrix}
		 		-12(\lambda^p + \lambda^{-p})		& 6p(\lambda^p - \lambda^{-p})	\\
		 		-6p(\lambda^p - \lambda^{-p})		& 2p^2(\lambda^p + \lambda^{-p})
		 	\end{bmatrix}
		 	\right) 
		\end{equation}
		Using Euler's identity $\lambda = e^{i\kappa}$, the above expression becomes:
		\begin{equation}
		 \sum_{p=1}^P \mathcal{K}_p(\lambda, \beta_p) =
		 \begin{bmatrix}
		 	\alpha			& \gamma\\
		 	-\gamma			& \delta
		 \end{bmatrix}
		\end{equation}
		where the scalar coefficients are given by: 
		\begin{align}
		 	\alpha 	&= 24 	\sum_{p=1}^P \beta_p[1 - \cos(\kappa p)]/p^3		\\
		 	\delta 	&= 4 	\sum_{p=1}^P \beta_p [ 2 + \cos(\kappa p)]/p	\\
		 	\gamma	&= 12i	\sum_{p=1}^P \beta_p \sin(\kappa p)/p^2
		\end{align}
		The dispersion relation follows from the determinant condition:
		\begin{equation}
		 	\sum_{p=1}^P \mathcal{K}_p(\lambda, \beta_p) - \omega^2 \mathbf{M}
		\end{equation}
		where the mass matrix $\mathbf{M}$ is defined in Section~\ref{section:Euler-BernoulliBeamLattices}.  
		A straightforward computation yields:
		\begin{equation}
			(\alpha\delta + \gamma^2) - (\mu\alpha + \delta)\omega^2 + \mu\omega^4 = 0
		\end{equation}
		This quartic in $\omega$ reduces to a quadratic in $\omega^2$, whose two positive roots $\omega_- \leq \omega_+$ are:
		\begin{equation}
			\omega_\pm^2(\lambda, \beta) = \left[ \mu\alpha + \delta \pm \sqrt{(\mu\alpha - \delta)^2 - 4\mu\gamma^2} \right]/2\mu
		\end{equation}
		In particular, in the case $P=1$, the dispersion branches $\omega_\pm$ are injective on $[0,\pi]$, as they reduce to sums of injective functions.
		It is useful to note that $\omega_\pm^2$ can be decomposed into a linear and a nonlinear part with respect to the stiffness coefficients $\beta$:
		\begin{equation}
			\omega_\pm^2(\lambda, \beta) = \mathcal{L}(\lambda, \beta) \pm G(\lambda, \beta)
		\end{equation}
		with
		\begin{align}
			\mathcal{L}(\lambda, \beta) &:= (\mu\alpha + \delta)/2\mu, \\
			G(\lambda, \beta) &:= \sqrt{(\mu\alpha - \delta)^2 - 4\mu\gamma^2}/2\mu.
		\end{align}
		If, for a fixed \(\lambda_i\), one prescribes \(\omega_-(\lambda_i, \beta) = a\) and \(\omega_+(\lambda_i, \beta) = b\),  
		the interpolation system~\eqref{eq:system_interpolation} simplifies to:
		\begin{equation}
			\begin{cases}
				\mathcal{L}(\lambda_i, \beta) 	&= (b + a)/2		\\
				G(\lambda_i, \beta)				&= (b - a)/2
			\end{cases}
		\end{equation}

		\subsubsection{Existence and uniqueness in dispersion curve for the Euler–Bernoulli beam lattice}
		\label{sec:other_dispersion_curves}

		\added[id = RB]{This section presents several parameterizations of dispersion curves for the beam-interaction nonlocal model,
		in order to further illustrate the discussion on the existence and uniqueness of solutions to system~\ref{eq:system_interpolation}.}

\begin{figure}[ht]
    		\centering
    		\begin{subfigure}{0.30\textwidth}
        		\centering
        		\includegraphics[width=\linewidth]{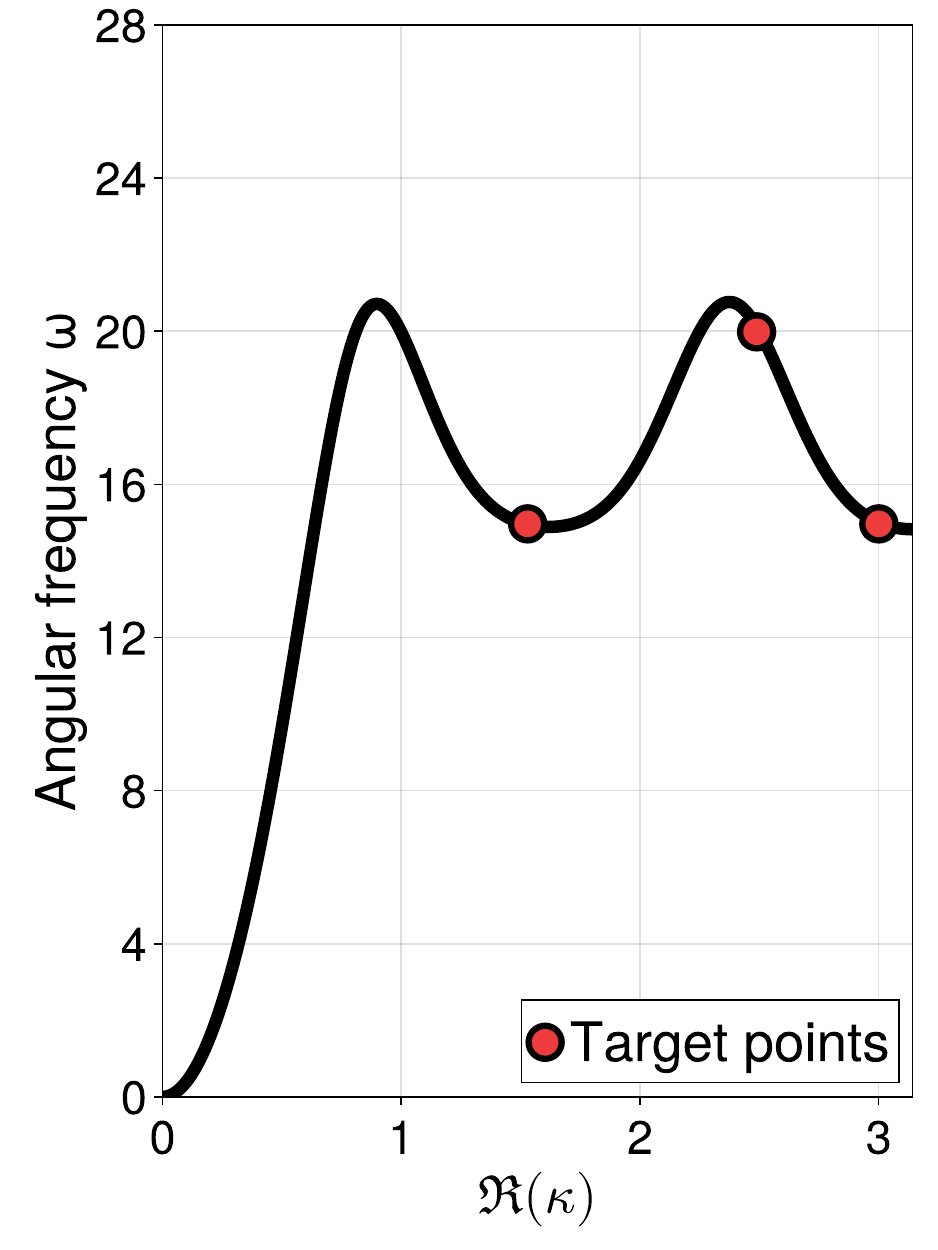}
        		\caption{}
				\label{fig:dispersion_curves_04}
    		\end{subfigure}
			\hspace{0.02\textwidth} 
    		\begin{subfigure}{0.30\textwidth}
        		\centering
        		\includegraphics[width=\linewidth]{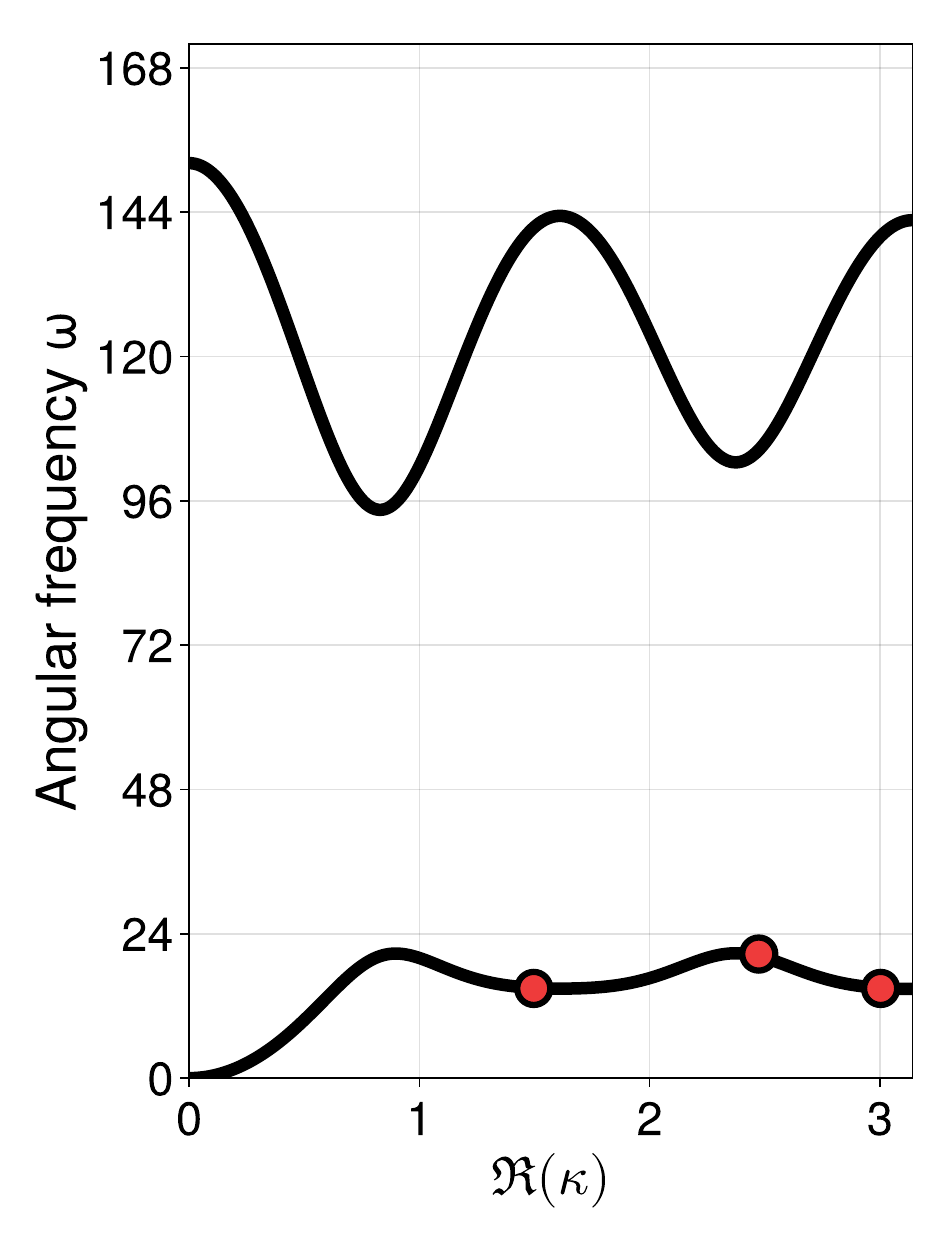}
        		\caption{}
				\label{fig:dispersion_curves_05}
    		\end{subfigure}
			\caption{\added[id = RB]{(\subref{fig:dispersion_curves_04}) Parameterization of a fourth-order nonlocal model featuring oscillations on the $\omega_-$ branch. 
			Only three interpolation points are used (see Table~\ref{table:dispersion_curves_annex_01}), making system~\ref{eq:system_interpolation} underdetermined. 
			This nonetheless allows at least one solution with strictly positive stiffnesses, which is not achievable with a third-order nonlocal model.
			(\subref{fig:dispersion_curves_05}) Same model as in (\subref{fig:dispersion_curves_04}), now displaying the $\omega_+$ branch. 
			Imposing oscillations on $\omega_-$ induces much stronger oscillations on the $\omega_+$ branch.}}
			\label{fig:dispersion_curves_annex_01}
		\end{figure}

\begin{figure}[ht]
    		\centering
    		\begin{subfigure}{0.30\textwidth}
        		\centering
        		\includegraphics[width=\linewidth]{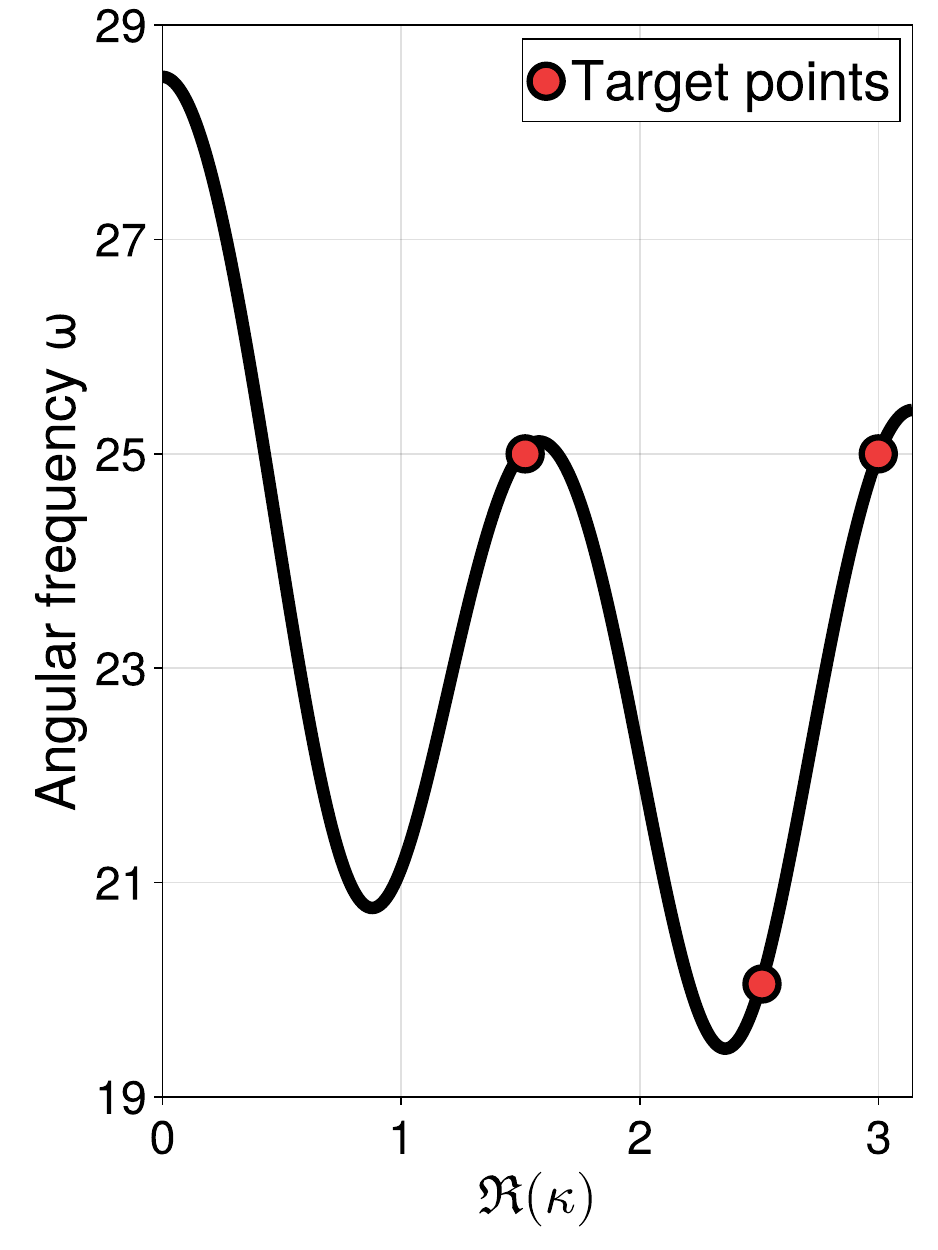}
        		\caption{}
				\label{fig:dispersion_curves_06}
    		\end{subfigure}
			\hspace{0.02\textwidth} 
    		\begin{subfigure}{0.30\textwidth}
        		\centering
        		\includegraphics[width=\linewidth]{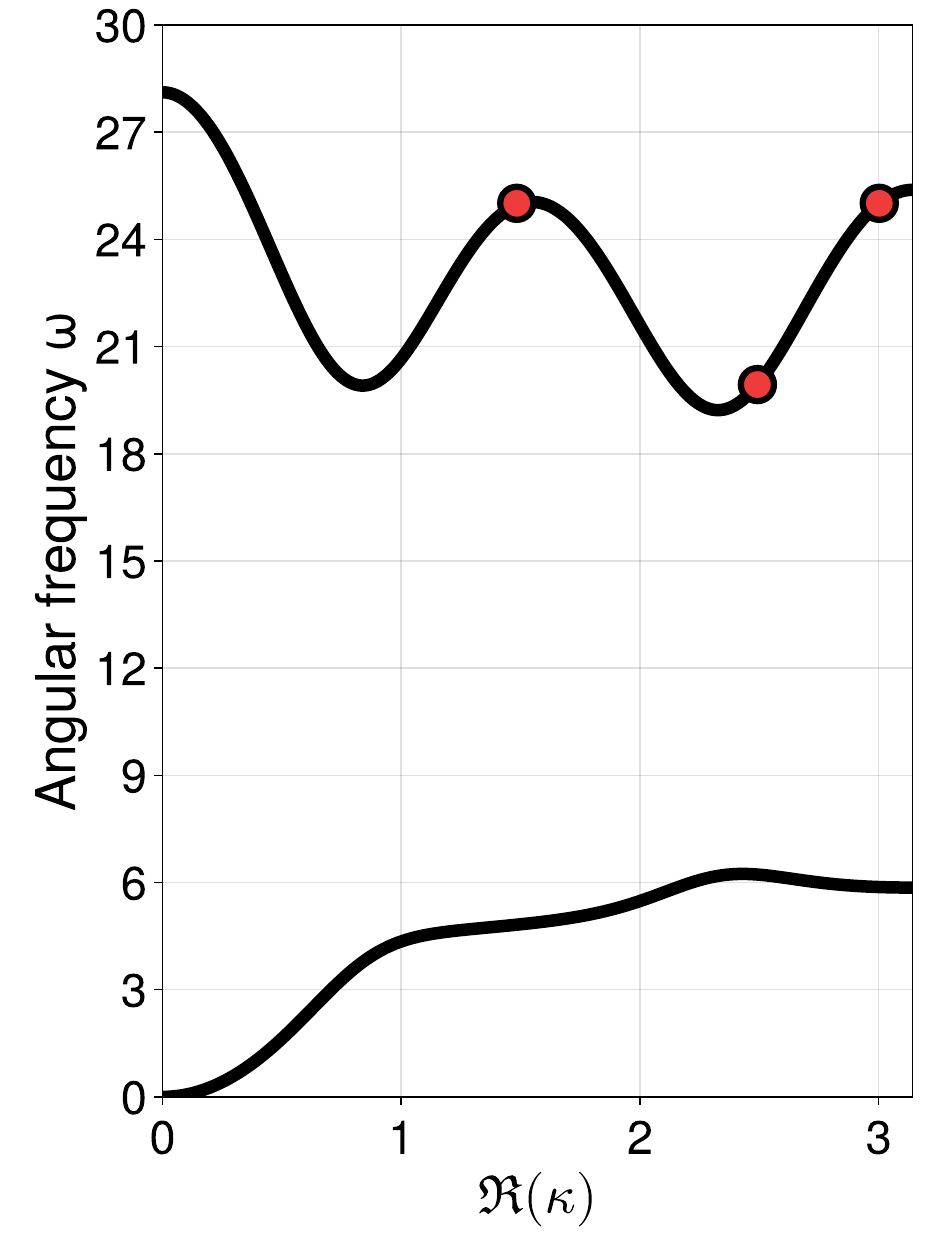}
        		\caption{}
				\label{fig:dispersion_curves_07}
    		\end{subfigure}
			\hspace{0.02\textwidth} 
			\begin{subfigure}{0.30\textwidth}
        		\centering
        		\includegraphics[width=\linewidth]{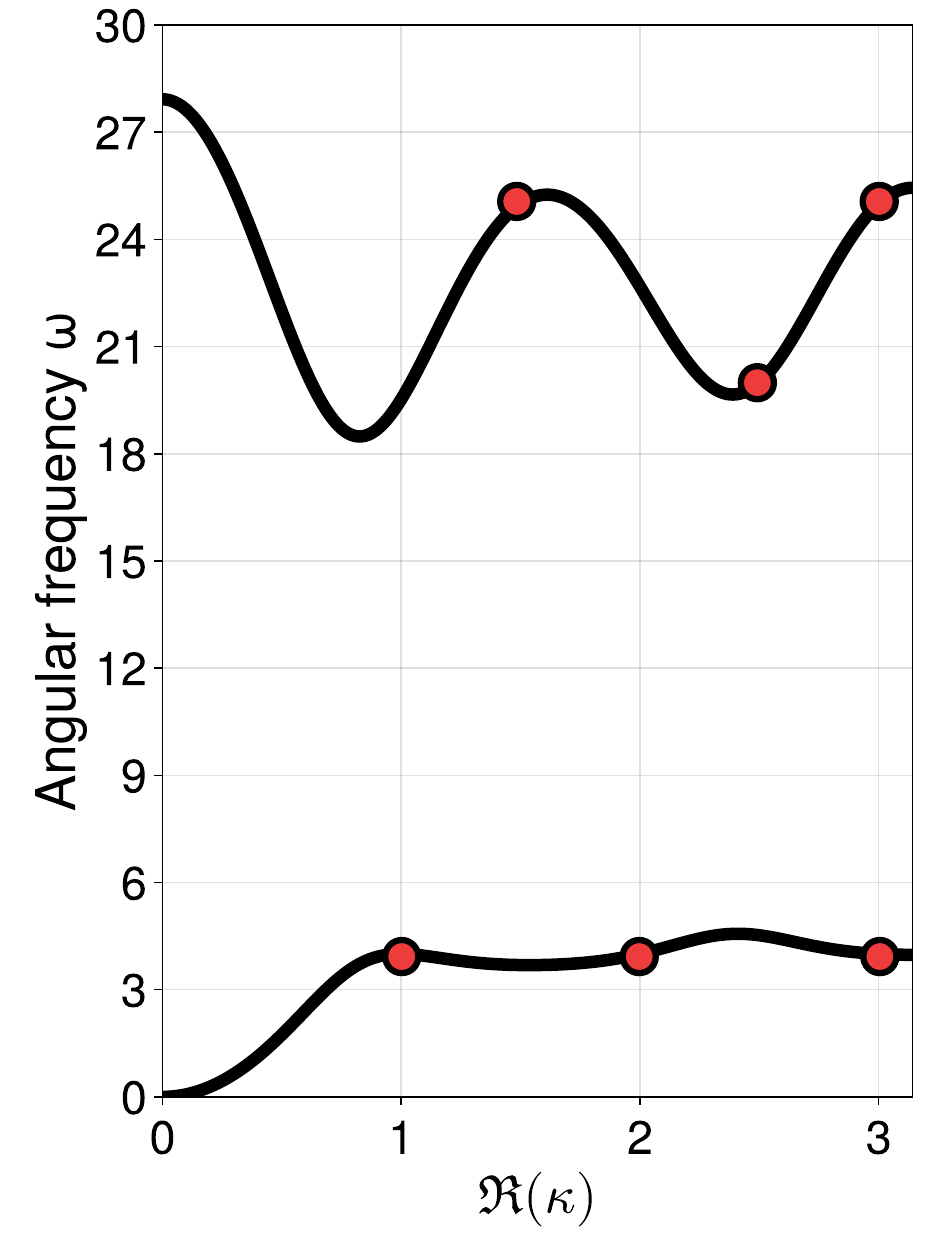}
        		\caption{}
				\label{fig:dispersion_curves_08}
    		\end{subfigure}
			\caption{\added[id = RB]{(\subref{fig:dispersion_curves_06}) Parameterization of a fourth-order nonlocal model with oscillations on the $\omega_+$ branch. 
			As in Example~\ref{fig:dispersion_curves_annex_01}, only three interpolation points are used (see Table~\ref{table:dispersion_curves_annex_02}),
			making system~\ref{eq:system_interpolation} underdetermined. 
			(\subref{fig:dispersion_curves_07}) Same model as in (\subref{fig:dispersion_curves_06}), now displaying the $\omega_-$ branch. 
			(\subref{fig:dispersion_curves_08}) Parameterization of a sixth-order nonlocal model using the same interpolation points as in (\subref{fig:dispersion_curves_06})
			on $\omega_+$, complemented with additional interpolation points (see Table~\ref{table:dispersion_curves_annex_02}) on $\omega_-$.}}
			\label{fig:dispersion_curves_annex_02}
		\end{figure}

		\added[id = RB]{Figures~\ref{fig:dispersion_curves_04} and~\ref{fig:dispersion_curves_06} first consider cases where the branches $\omega_-$ and $\omega_+$
		are independently parameterized to exhibit strong oscillations. In both examples, three interpolation points are imposed.
		To obtain positive solutions of system~\ref{eq:system_interpolation}, the problem must be made underdetermined by considering fourth-order nonlocal models.
		Figure~\ref{fig:dispersion_curves_annex_01} shows that oscillations introduced on the $\omega_-$ branch induce oscillations of much larger amplitude on the $\omega_+$ branch.
		In this situation, adding interpolation points on $\omega_+$ in an attempt to flatten the branch leads to the absence of solutions.
		Conversely, Figure~\ref{fig:dispersion_curves_08} shows that when strong oscillations are prescribed on $\omega_+$,
		the branch $\omega_-$ can still be parameterized with relatively small oscillations.}

\begin{figure}[ht]
    		\centering
    		\begin{subfigure}{0.30\textwidth}
        		\centering
        		\includegraphics[width=\linewidth]{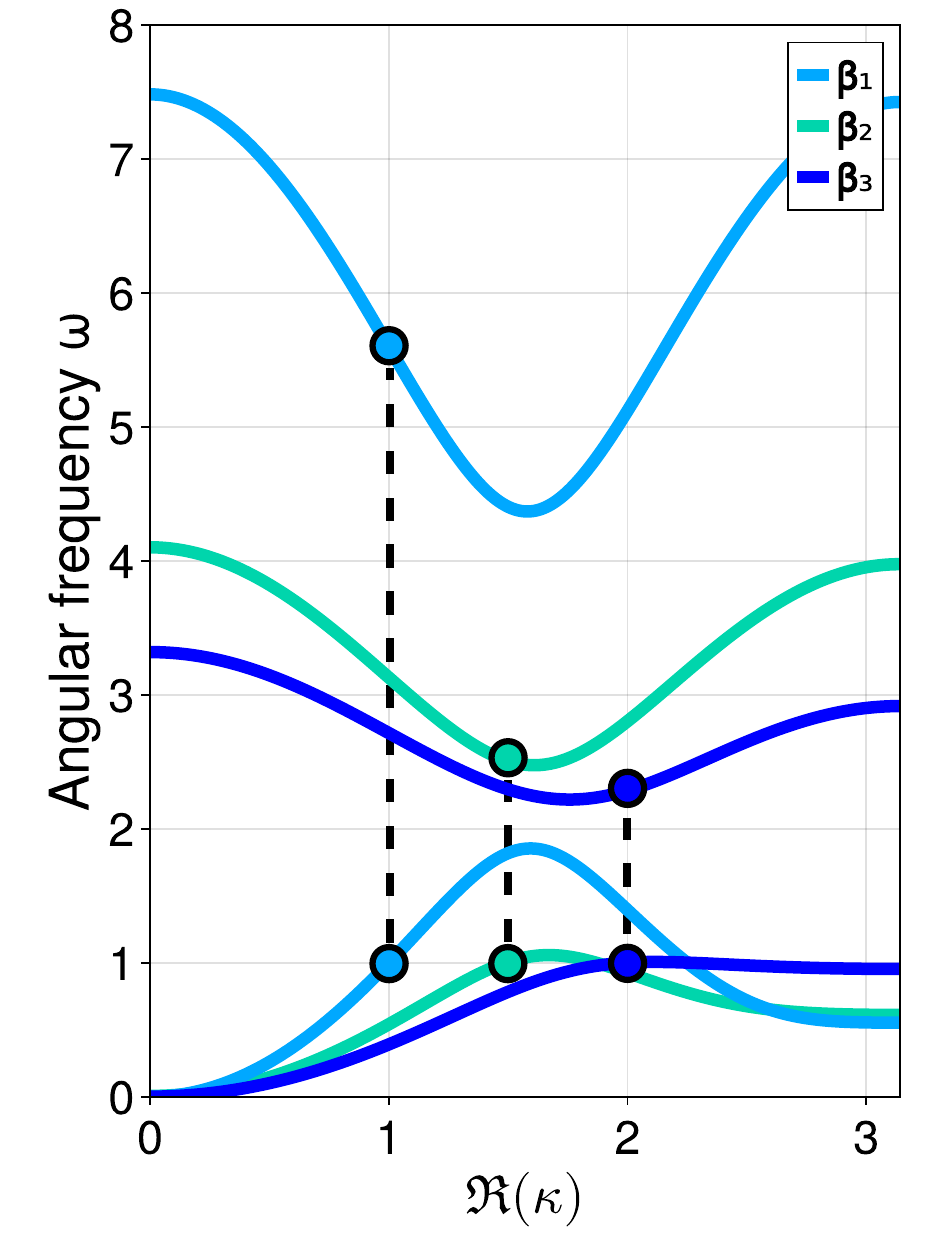}
        		\caption{}
				\label{fig:dispersion_curves_09}
    		\end{subfigure}
			\hspace{0.02\textwidth} 
    		\begin{subfigure}{0.30\textwidth}
        		\centering
        		\includegraphics[width=\linewidth]{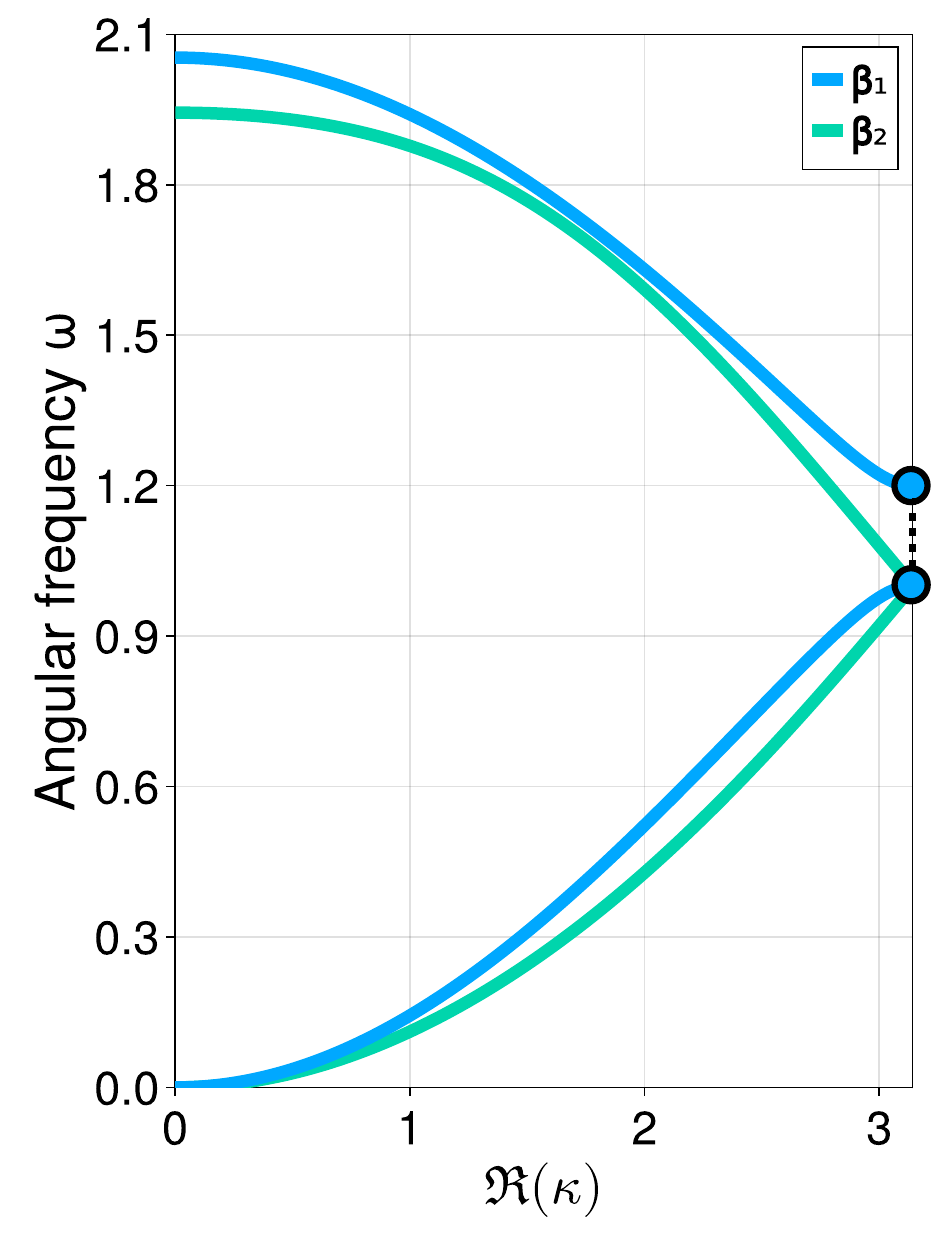}
        		\caption{}
				\label{fig:dispersion_curves_10}
    		\end{subfigure}
			\caption{\added[id = RB]{(\subref{fig:dispersion_curves_09}) Three examples of parametrizations of the second-order nonlocal model.
			The objective is to construct dispersion curves with two interpolation points located on the branches $\omega_-$ and $\omega_+$,
			respectively, associated with the same wavenumber and angular frequencies as close as possible.
			For a fixed interpolation point $(\kappa_0, \omega_0)$ on $\omega_-$,
			the corresponding point $(\kappa_0, \omega_{\text{min}})$ is determined such that system~\ref{eq:system_interpolation}
			admits a positive solution (see Table~\ref{table:dispersion_curves_annex_03a}).
			As $\kappa_0$ increases, $\omega_{\text{min}}$ can be selected closer to $\omega_0$.
			The critical case is reached in Fig.~(\subref{fig:dispersion_curves_10}), where $\kappa_0 = \pi$.
			In particular, if the positivity constraint on the stiffness parameters is relaxed,
			a model can be designed such that $\omega_0 = \omega_-$ (see Table~\ref{table:dispersion_curves_annex_03b}).}}
			\label{fig:dispersion_curves_annex_03}
		\end{figure}

		\added[id = RB]{
		More generally, the behavior of the two branches is strongly coupled: prescribing one branch reduces the flexibility available for parameterizing the other.
		This effect is illustrated in Figure~\ref{fig:dispersion_curves_annex_03},
		where two interpolation points are chosen as close as possible on the branches $\omega_-$ and $\omega_+$.
		A critical proximity distance is observed, below which no solution to system~\ref{eq:system_interpolation} can be found.}

\begin{figure}[ht]
    		\centering
    		\begin{subfigure}{0.30\textwidth}
        		\centering
        		\includegraphics[width=\linewidth]{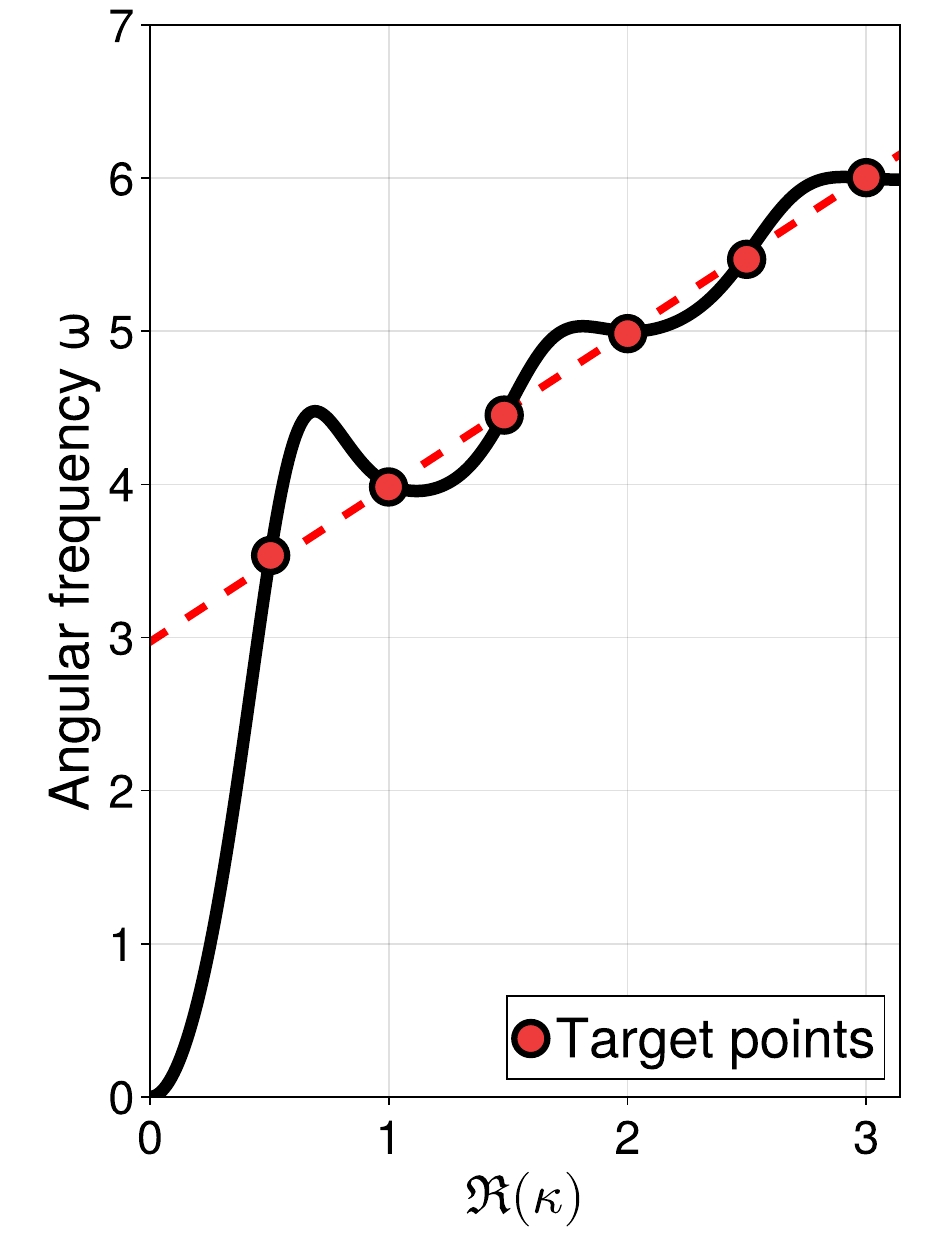}
        		\caption{}
				\label{fig:dispersion_curves_11}
    		\end{subfigure}
			\hspace{0.02\textwidth} 
    		\begin{subfigure}{0.30\textwidth}
        		\centering
        		\includegraphics[width=\linewidth]{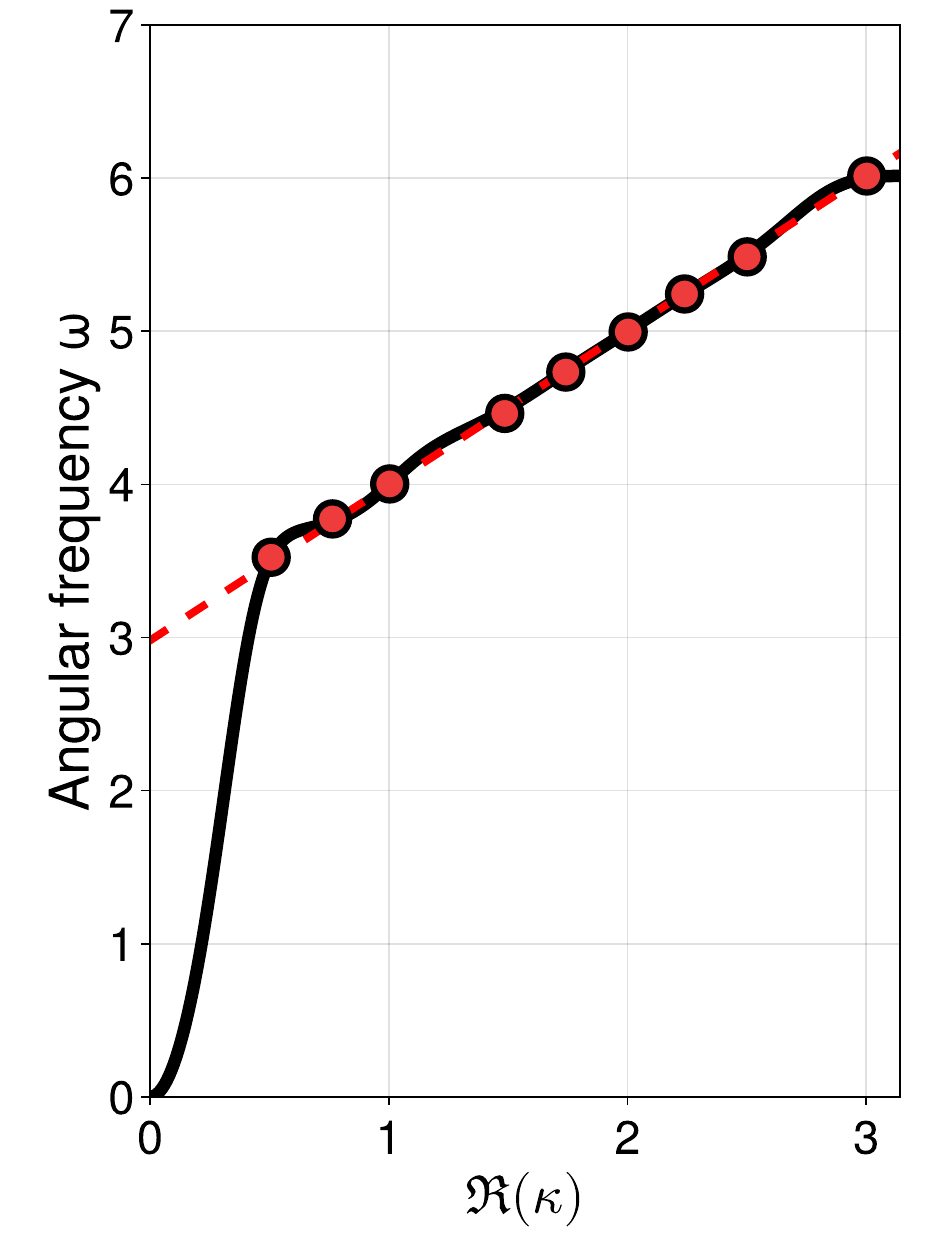}
        		\caption{}
				\label{fig:dispersion_curves_12}
    		\end{subfigure}
			\hspace{0.02\textwidth} 
    		\begin{subfigure}{0.30\textwidth}
        		\centering
        		\includegraphics[width=\linewidth]{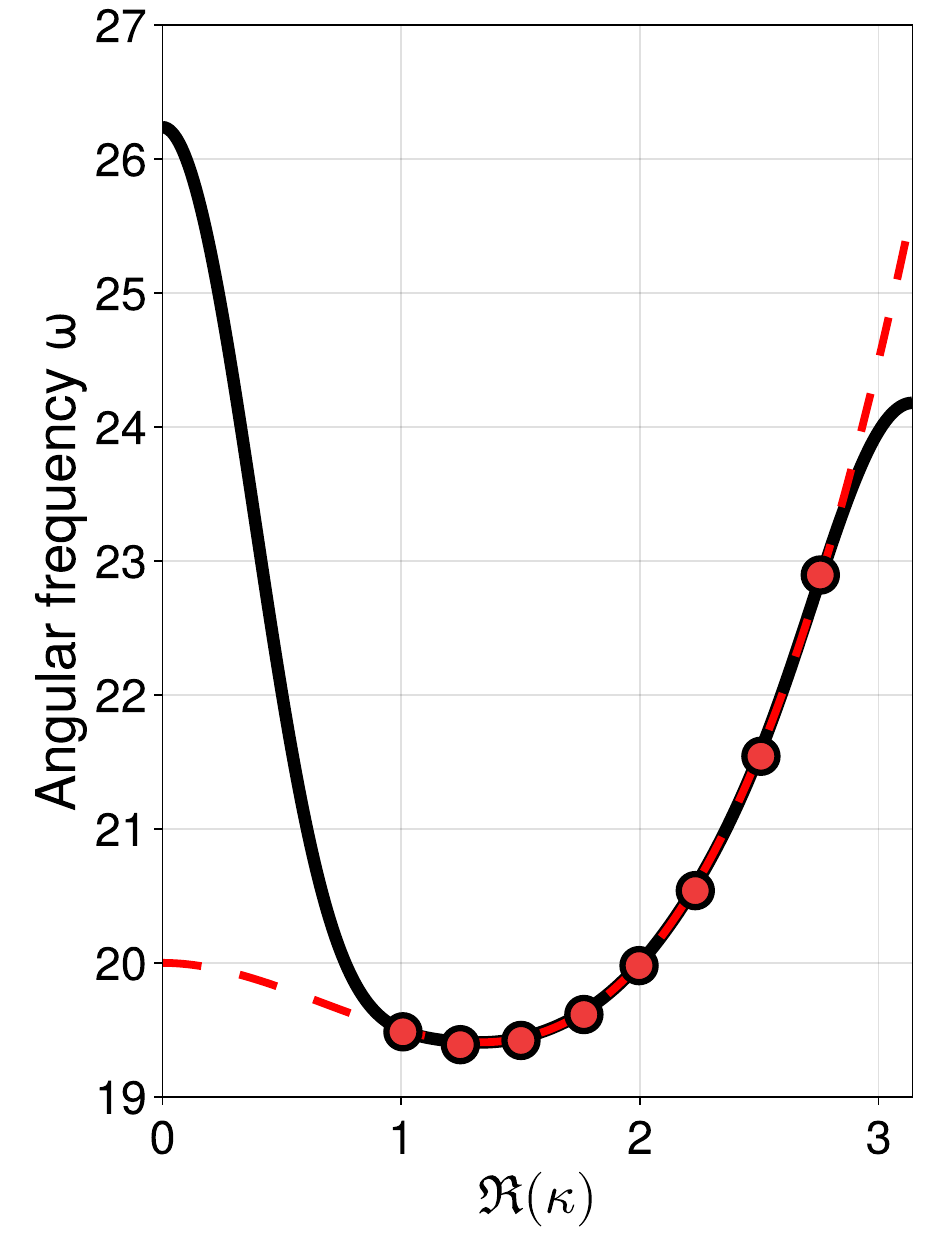}
        		\caption{}
				\label{fig:dispersion_curves_13}
    		\end{subfigure}
			\caption{\added[id = RB]{Example of model parametrizations designed to approximate a prescribed target curve.
			(\subref{fig:dispersion_curves_11}) The branch $\omega_-$ is parametrized so as to obtain a straight line of slope $1$ over the interval $(\kappa = 0.5, \kappa = 3.0)$.
			(\subref{fig:dispersion_curves_12}) Same objective as in (\subref{fig:dispersion_curves_11}),
			using a larger number of interpolation points (see Table~\ref{table:dispersion_curves_annex_04}).
			(\subref{fig:dispersion_curves_13}) Parametrization of the branch $\omega_+$
			to approximate the target curve analytically defined by $\omega = 20 - \kappa^2 + 0.5\kappa^3$
			over the interval $(\kappa = 1.0, \kappa = 3.0)$ (see Table~\ref{table:dispersion_curves_annex_04}).}}
			\label{fig:dispersion_curves_annex_04}
		\end{figure}

		\added[id = RB]{
		Figure~\ref{fig:dispersion_curves_annex_04} presents two examples of more global customization of the dispersion curve.
		In both cases, example~\ref{fig:dispersion_curves_11}, where the branch $\omega_-$ is prescribed, and example~\ref{fig:dispersion_curves_13},
		where $\omega_+$ is prescribed, a portion of the first Brillouin zone must remain unconstrained in order for a solution to exist.}

\begin{figure}[ht]
    		\centering
    		\begin{subfigure}{0.30\textwidth}
        		\centering
        		\includegraphics[width=\linewidth]{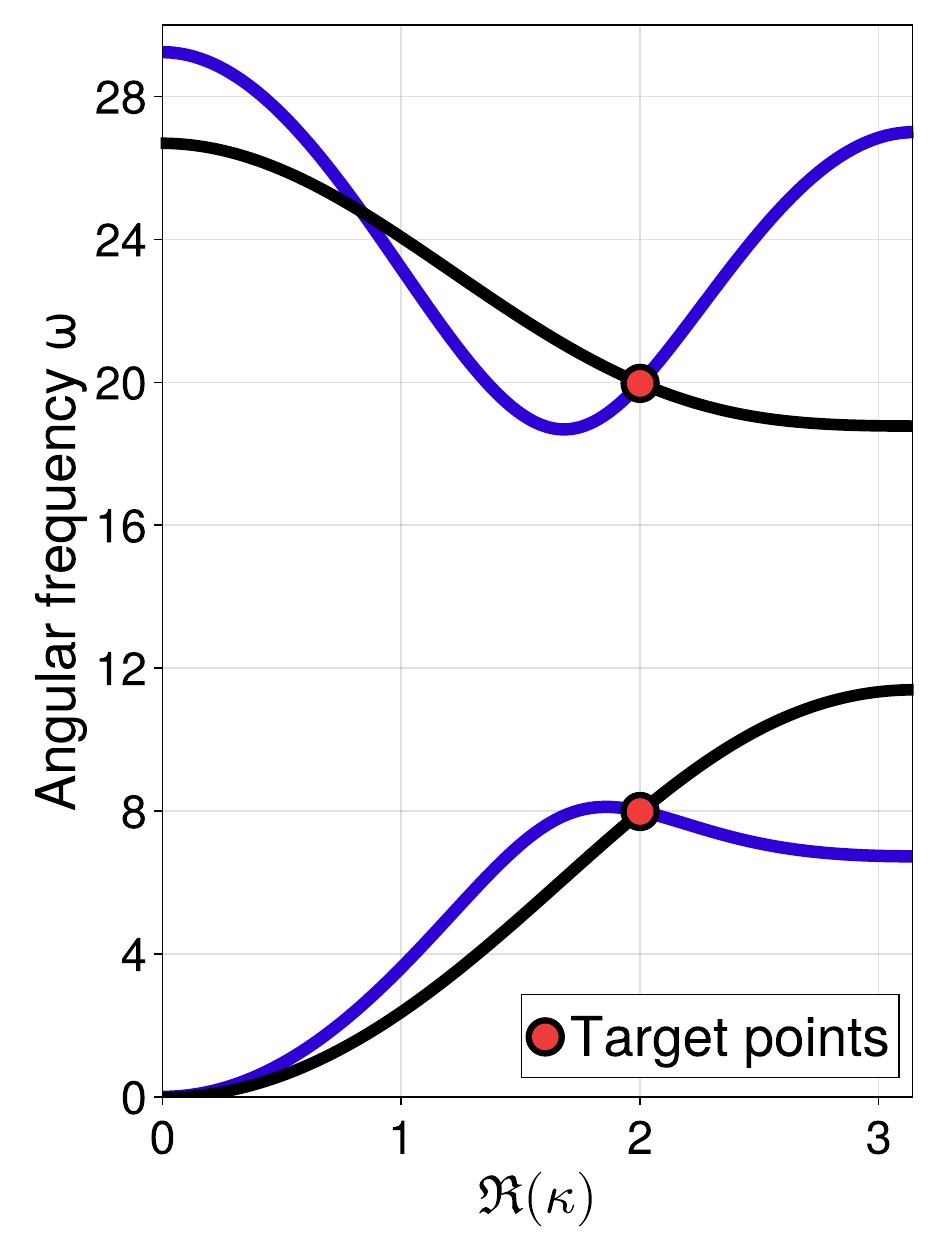}
        		\caption{}
				\label{fig:dispersion_curves_14}
    		\end{subfigure}
			\hspace{0.02\textwidth} 
    		\begin{subfigure}{0.30\textwidth}
        		\centering
        		\includegraphics[width=\linewidth]{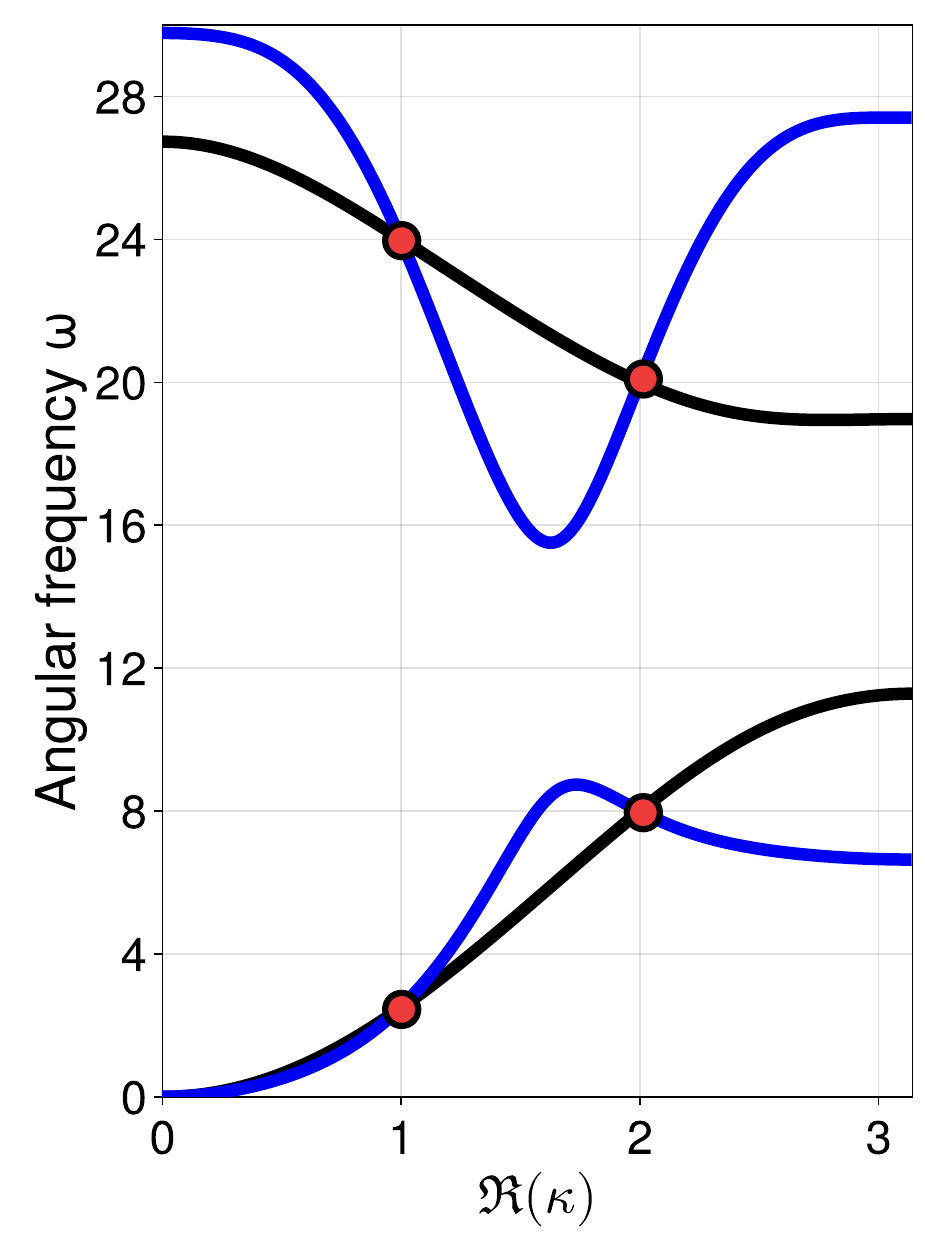}
        		\caption{}
				\label{fig:dispersion_curves_15}
    		\end{subfigure}
			\caption{\added[id = RB]{(\subref{fig:dispersion_curves_14}) Example of a parametrization of a second-order nonlocal model using two interpolation points.
			For this set of interpolation points, system \ref{eq:system_interpolation} admits two solutions with positive stiffness parameters (see Table~\ref{table:dispersion_curves_annex_05}).
			(\subref{fig:dispersion_curves_14}) Example of a parametrization of a fourth-order nonlocal model using four interpolation points.
			For this set of interpolation points, system \ref{eq:system_interpolation} admits two solutions.
			However, only the one shown in black has strictly positive stiffness parameters (see Table~\ref{table:dispersion_curves_annex_05}).}}
			\label{fig:dispersion_curves_annex_05}
		\end{figure}

		\added[id = RB]{
		Finally, Figures~\ref{fig:dispersion_curves_annex_05} illustrate two distinct cases in which nonlocal models of the same order produce dispersion
		curves passing through the same interpolation points. In other words, system~\ref{eq:system_interpolation} does not admit a unique solution.
		In such situations, additional selection criteria may be introduced,
		such as the positivity of stiffness coefficients or the enforcement of a prescribed behavior between interpolation points.}

		\clearpage
		\subsection{Calculation of stiffnesses}

		\subsubsection{Positivity constraint on stiffness parameters}
		\label{sec:positivity_constraint}

		In the case of beam-type interactions, the interpolation system~\eqref{eq:system_interpolation} is nonlinear.  
		To solve it, iterative numerical schemes of the Newton type are employed.  
		Throughout this work, an important design constraint has been systematically enforced: all presented models are obtained using strictly positive stiffness values.  
		This choice ensures the physical realism of the constructed lattices and avoids configurations that would correspond to unstable or non-physical mechanical responses.  

		Accordingly, a positivity constraint is incorporated into the numerical solvers.  
		In contrast, for the special case of nonlocal spring-interaction lattices discussed in Section~\ref{part:monoatomic-lattice},  
		the system~\eqref{eq:system_interpolation} is linear.  
		In such cases, a direct solution always exists for any set of prescribed target points, yielding stiffness values that reproduce the desired dispersion curve.  
		However, negative stiffnesses may arise in these unconstrained solutions.  
		To prevent this, the system is preferably solved using a constrained optimization approach,
		such as the Non-Negative Least Squares (NNLS) method developed by Lawson and Hanson~\cite{lawson1995solving}.  
		This formulation guarantees physically admissible, strictly positive stiffness distributions while maintaining accurate interpolation of the prescribed dispersion points.

		\subsubsection{Approximation of target dispersion profiles}

		In some cases, it may be desirable to tune a nonlocal lattice model so that its dispersion profile approximates a predefined target curve.  
		To illustrate this, the interpolation method is applied here to nonlocal spring-interaction models.  
		This scenario has been investigated in~\cite{kazemi2023non}, three target dispersion profiles from that study are revisited using the present approach.

		Without enforcing stiffness positivity, the proposed interpolation method yields models whose dispersion curves closely match the target profiles.  
		However, when the positivity constraint on the stiffnesses is imposed, globally accurate reconstructions become infeasible.  
		In such situations, it is necessary to restrict attention to a specific region of interest along the dispersion curve and focus the interpolation within that interval.

		Initially, interpolation points are chosen empirically to provide a reasonable match to the overall target, denoted as case \textbf{WN} (without constraint).  
		Although this approach effectively reproduces the qualitative shape of the curve, it may yield negative stiffness values.  
		To address this issue, the same set of interpolation points is used within a Non-Negative Least Squares (NNLS) formulation to enforce stiffness positivity,
		yielding the case \textbf{OP~1}.  
		This procedure guarantees physical admissibility but generally reduces the global accuracy of the reconstruction.

		To improve the trade-off between fidelity and feasibility, a localized optimization strategy is introduced.  
		A region of interest is selected on the target curve, and new interpolation points are chosen within that interval.  
		The NNLS regression is then reapplied, yielding the case \textbf{OP~2}.  
		This local parametrization improves the agreement between the model and the target curve in the selected frequency range, while maintaining positive stiffnesses throughout.

\begin{figure}[ht]
    		\centering

    		\begin{subfigure}[b]{\textwidth}
        		\centering
       			 \begin{subfigure}[b]{0.37\textwidth}
            		\includegraphics[width=\textwidth]{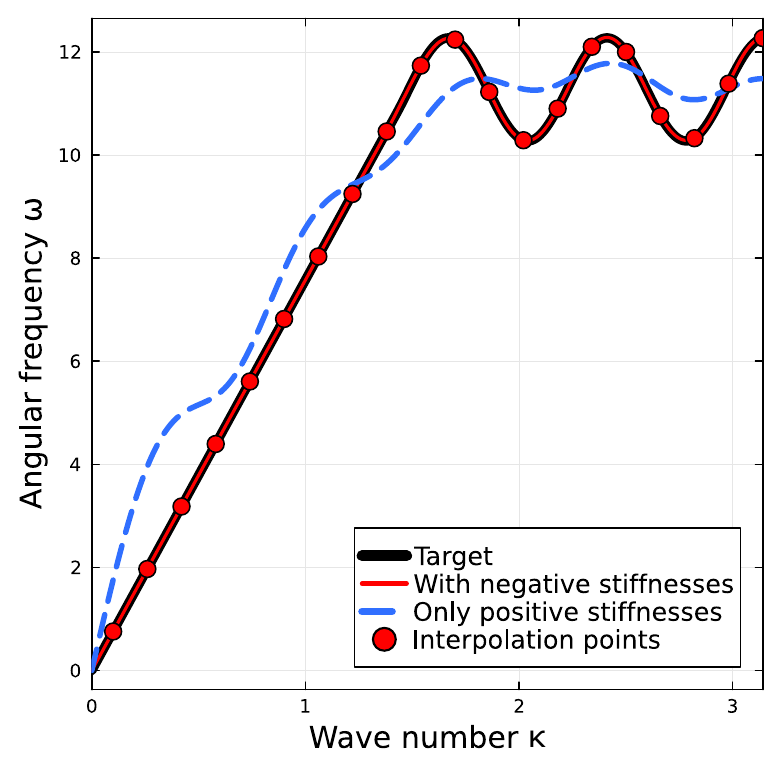}
        		\end{subfigure}
        		\hspace{0.02\textwidth} 
        		\begin{subfigure}[b]{0.37\textwidth}
            		\includegraphics[width=\textwidth]{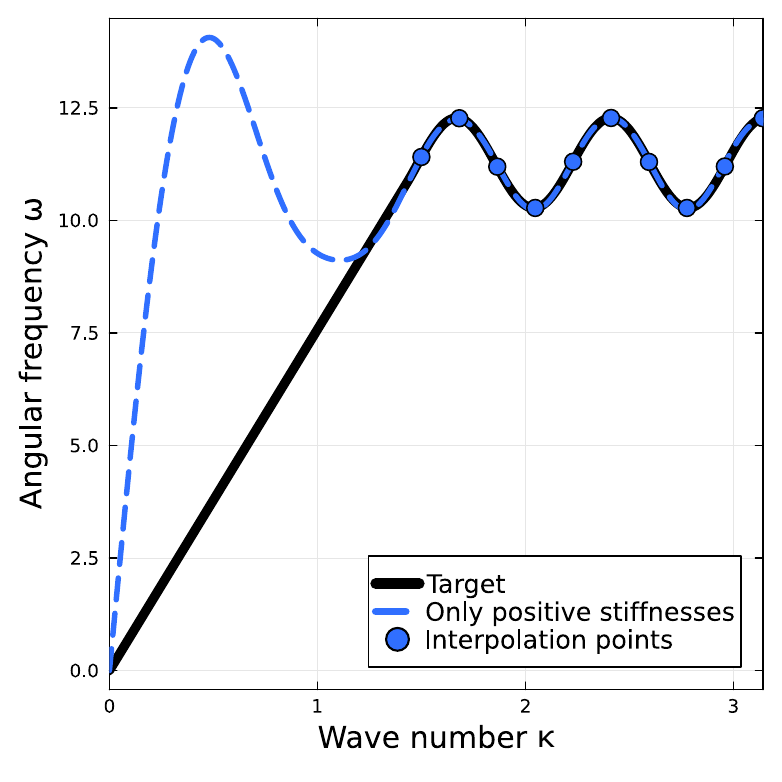}
        		\end{subfigure}
        		\caption{}
    		\end{subfigure}

    		\vskip\baselineskip

    		\begin{subfigure}[b]{\textwidth}
        		\centering
        		\begin{subfigure}[b]{0.37\textwidth}
            		\includegraphics[width=\textwidth]{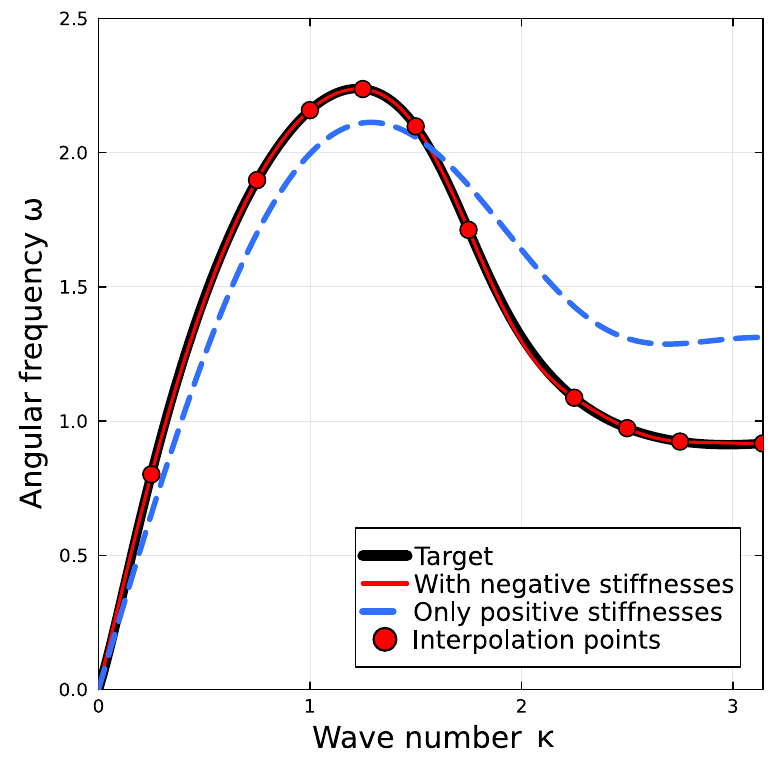}
        		\end{subfigure}
				\hspace{0.02\textwidth} 
        		\begin{subfigure}[b]{0.37\textwidth}
            		\includegraphics[width=\textwidth]{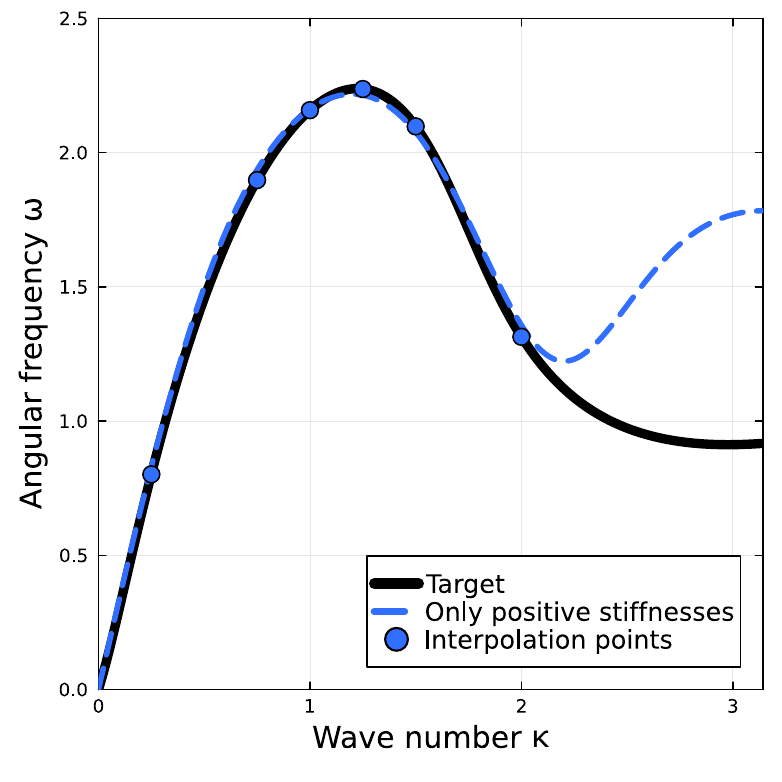}
        		\end{subfigure}
        		\caption{}
    		\end{subfigure}

    		\vskip\baselineskip

    		\begin{subfigure}[b]{\textwidth}
        		\centering
        		\begin{subfigure}[b]{0.37\textwidth}
            		\includegraphics[width=\textwidth]{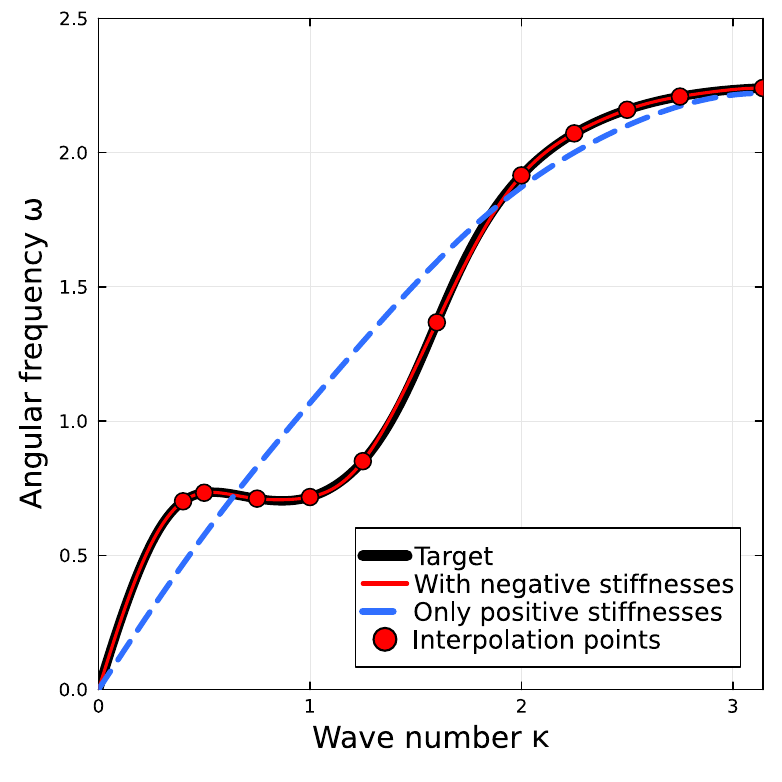}
        		\end{subfigure}
        		\hspace{0.02\textwidth} 
        		\begin{subfigure}[b]{0.37\textwidth}
            		\includegraphics[width=\textwidth]{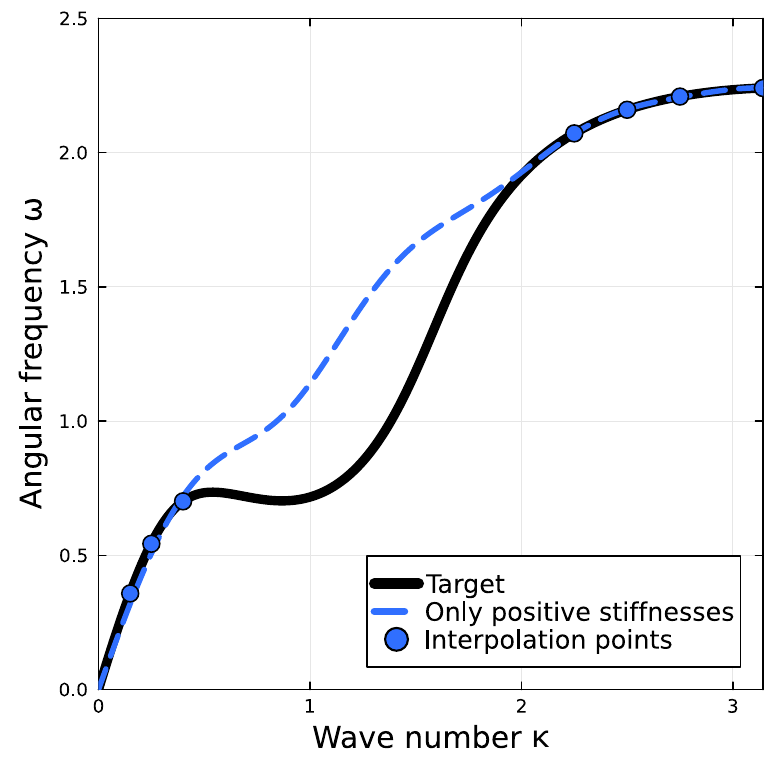}
        		\end{subfigure}
        		\caption{}
    		\end{subfigure}

    		\caption{Comparison between interpolation strategies.  
			The target dispersion curve is shown in black.
			Red markers correspond to global interpolation points (WN), while blue markers indicate locally optimized points (OP~2) within the selected region of interest.}
    		\label{fig:comparaison_dispersions}
		\end{figure}

		The different results obtained under the three interpolation strategies, unconstrained (WN),
		globally constrained (OP~1), and locally constrained (OP~2), are summarized in Table~\ref{table:GVD} and illustrated in Figure~\ref{fig:comparaison_dispersions},
		highlighting the trade-off between accuracy and physical admissibility.

		\clearpage
		\subsubsection{Numerical parameters used in the figures}
		\label{sec:Table_values}
		The following tables summarize the numerical parameters employed in the simulations associated with each figure.  
		They include the prescribed interpolation points $(\kappa_i, \omega_i)$ and the corresponding optimized stiffness values $\beta_p$.

		\begin{table}[ht]
			\centering
			\begin{tabular}{|l|l|l||l|l|l|}
				\hline
				\multicolumn{6}{|c|}{\text{Parameters of figure \ref{fig:dispersion_beam}}} \\
				\hline
				\multicolumn{3}{|c||}{$\qquad$ \textbf{(a)} $\qquad$ } & \multicolumn{3}{c|}{$\qquad$ \textbf{(b)}$\qquad$ }	\\
				\hline
				$\kappa_1=1.5$	& $\omega_{1-}=2.0$	& $\beta_1= 0.0668$	& $\kappa_1=1.0$	& $\omega_{1+} = 6.0$	& $\beta_1=0.1078$	\\
				$\kappa_2=2.0$	& $\omega_{2-}=2.0$	& $\beta_2= 0.3490$	& $\kappa_2=1.5$	& $\omega_{2+} = 6.0$	& $\beta_2=0.1740$ \\
				$\kappa_3=3.0$	& $\omega_{3-}=2.0$	& $\beta_3= 0.4526$	& $\kappa_3=2.0$	& $\omega_{3+} = 6.0$	& $\beta_3=0.1780$	\\
								&					&					& $\kappa_4=3.0$	& $\omega_{4+} = 6.0$	& $\beta_4=0.1533$	\\
				\hline
				\multicolumn{3}{|c||}{$\qquad$ \textbf{(c)} $\qquad$ } &	\multicolumn{3}{|c|}{}	\\
				\hline
				$\kappa_1=1.5$	& $\omega_{1-}=1.0$	& $\beta_1=0.0116$	& \multicolumn{3}{|c|}{}	\\
				$\kappa_2=2.0$	& $\omega_{2-}=1.0$ & $\beta_2=0.0768$	& \multicolumn{3}{|c|}{}	\\
				$\kappa_3=3.0$	& $\omega_{3-}=1.0$ & $\beta_3=0.1790$	& \multicolumn{3}{|c|}{}	\\
				$\kappa_4=1.5$	& $\omega_{4+}=6.0$	& $\beta_4=0.2347$	& \multicolumn{3}{|c|}{}	\\
				$\kappa_5=2.0$	& $\omega_{5+}=6.0$	& $\beta_5=0.3316$	& \multicolumn{3}{|c|}{}	\\
				$\kappa_6=3.0$	& $\omega_{6+}=6.0$	& $\beta_6=0.2269$	& \multicolumn{3}{|c|}{}	\\	
				\hline		
			\end{tabular}
			\caption{Examples of parameterization.}
			\label{table:dispersion_beam}
		\end{table}

		\begin{table}[ht]
			\centering
			\begin{tabular}{|l|l|l||l|l|l|}
				\hline
				\multicolumn{6}{|c|}{\text{Parameters of figure \ref{fig:rotons_dispersion}}} \\
				\hline
				\multicolumn{3}{|c||}{$\qquad$ \textbf{(a)} $\qquad$ } & \multicolumn{3}{c|}{$\qquad$ \textbf{(b)}$\qquad$ }	\\
				\hline
				$\kappa_1=1.8$	& $\omega_{1+}=7.9$	& $\beta_1=0.2337$	& $\kappa_1=1.0$	& $\omega_{1-} = 6.0$	& $\beta_1=0.75$	\\
				$\kappa_2=2.0$	& $\omega_{2+}=8.0$	& $\beta_2=0.0908$	& $\kappa_2=\pi$	& $\omega_{2-} = 6.0$	& $\beta_2=19.3979$ \\
				$\kappa_3=2.2$	& $\omega_{3+}=7.9$	& $\beta_3=0.5148$	& 					& 						& 				\\
				\hline	
			\end{tabular}
			\caption{Parameterization of rotons.}
			\label{table:Rotons}
		\end{table}

		\begin{table}[ht]
			\centering	
			\begin{tabular}{|l|l|l||l|l|l|}
				\hline
				\multicolumn{6}{|c|}{\text{Parameters of figure \ref{fig:GVD_curve}}} \\
				\hline
				\multicolumn{3}{|c||}{$\qquad$ \textbf{(a)} $\qquad$ } & \multicolumn{3}{c|}{$\qquad$ \textbf{(b)}$\qquad$ }	\\
				\hline
				$\kappa_1=1.3$	& $\omega_{1+}=12.4$	& $\beta_1= 0.8601$	& $\kappa_1=1.3$	& $\omega_{1+} = 12.6$	& $\beta_1=0.4744$	\\
				$\kappa_2=1.5$	& $\omega_{2+}=12.0$	& $\beta_2= 0.2290$	& $\kappa_2=1.5$	& $\omega_{2+} = 12.0$	& $\beta_2=1.8977$ \\
				$\kappa_3=1.7$	& $\omega_{3+}=11.6$	& $\beta_3= 0.0821$	& $\kappa_3=1.7$	& $\omega_{3+} = 11.8$	& $\beta_3=0.2367$	\\
				\hline
				\multicolumn{3}{|c||}{$\qquad$ \textbf{(c)} $\qquad$ } &	\multicolumn{3}{|c|}{}	\\
				\hline
				$\kappa_1=1.3$	& $\omega_{1+}=12.8$	& $\beta_1=0.1937$	& \multicolumn{3}{|c|}{}	\\
				$\kappa_2=1.5$	& $\omega_{2+}=12.0$ 	& $\beta_2=3.2726$	& \multicolumn{3}{|c|}{}	\\
				$\kappa_3=1.7$	& $\omega_{3+}=12.0$ 	& $\beta_3=0.1344$	& \multicolumn{3}{|c|}{}	\\
				\hline
			\end{tabular}
			\caption{Parameterization of the dispersion velocity group.}
			\label{table:GVD}
		\end{table}
		
		\begin{table}[ht]
			\centering
			\begin{tabular}{|l|l|l||l|l|l|}
				\hline
				\multicolumn{6}{|c|}{\text{Parameters of figure \ref{fig:evanescent_dispersion}}} \\
				\hline
				\multicolumn{3}{|c||}{$\qquad$ \textbf{(a)} $\qquad$ } & \multicolumn{3}{c|}{$\qquad$ \textbf{(b)}$\qquad$ }	\\
				\hline
				$\kappa_1=\pi$		& $\omega_{1-}=6.0$		& $\beta_1=0.5855$	& $\kappa_1=1.0$		& $\omega_{1-}=6.0$		& $\beta_1=0.7330$	\\
				$\kappa_2=\pi$		& $\omega_{2+}=12.0$	& $\beta_2=0.0625$	& $\kappa_2=\pi$		& $\omega_{2+}=12.0$	& $\beta_2=0.8492$ 	\\
				$\kappa_3=\pi-0.2i$	& $\omega_{3-}=6.5$		& $\beta_3=4.4422$	& $\kappa_3=\pi-0.7i$	& $\omega_{3-}=6.5$		& $\beta_3=0.4595$	\\
				\hline	
			\end{tabular}
			\caption{Parameterization of evanescent waves.}
			\label{table:evanescent_wave}
		\end{table}

		\begin{table}[ht]
			\centering
			\begin{tabular}{|l|l|l|}
				\hline
				\multicolumn{3}{|c|}{\text{Parameters of figure \ref{fig:dispersion_curves_annex_01}}} \\
				\hline
				$\kappa_1=1.5$		& $\omega_{1-}=15.0$	& $\beta_1=2.5306$		\\
				$\kappa_2=2.5$		& $\omega_{2-}=20.0$	& $\beta_2=18.9689$		\\
				$\kappa_3=3.0$		& $\omega_{3-}=15.5$	& $\beta_3=55.3363$		\\
									&						& $\beta_4=341.1581$	\\
				\hline	
			\end{tabular}
			\caption{Example of parameterization.}
			\label{table:dispersion_curves_annex_01}
		\end{table}

		\begin{table}[ht]
			\centering
			\begin{tabular}{|l|l|l||l|l|l|}
				\hline
				\multicolumn{6}{|c|}{\text{Parameters of figure \ref{fig:dispersion_curves_annex_02}}} \\
				\hline
				\multicolumn{3}{|c||}{$\qquad$ \textbf{(a)} $\qquad$ } & \multicolumn{3}{c|}{$\qquad$ \textbf{(c)}$\qquad$ }	\\
				\hline
				$\kappa_1=1.5$		& $\omega_{1+}=25.0$	& $\beta_1=0.9272$	& $\kappa_1=1.0$		& $\omega_{1+}=25.0$		& $\beta_1=0.2539$	\\
				$\kappa_2=2.5$		& $\omega_{2+}=20.0$	& $\beta_2=1.5202$	& $\kappa_2=2.5$		& $\omega_{2+}=20.0$		& $\beta_2=0.9264$ 	\\
				$\kappa_3=3.0$		& $\omega_{3+}=25.0$	& $\beta_3=4.4422$	& $\kappa_3=3.0$		& $\omega_{3+}=25.0$		& $\beta_3=1.9395$	\\
									&						& $\beta_4=8.1957$	& $\kappa_4=1.0$		& $\omega_{1-}=4.0$			& $\beta_4=9.3006$	\\
									&						&					& $\kappa_5=2.0$		& $\omega_{2-}=4.0$			& $\beta_5=0.4534$	\\
									&						&					& $\kappa_6=3.0$		& $\omega_{3-}=4.0$			& $\beta_6=0.7101$	\\
				\hline	
			\end{tabular}
			\caption{Examples of parameterization.}
			\label{table:dispersion_curves_annex_02}
		\end{table}

		\begin{table}[ht]
			\centering	
			\begin{tabular}{|l|l|l||l|l|l|}
				\hline
				\multicolumn{6}{|c|}{\text{Parameters of figure \ref{fig:dispersion_curves_09}}} \\
				\hline
				\multicolumn{3}{|c||}{$\boldsymbol{\beta}_1$ \qquad} & \multicolumn{3}{c|}{$\boldsymbol{\beta}_2$ \qquad}	\\
				\hline
				$\kappa_1=1.0$	& $\omega_{1-}=1.0$	& $\beta_1= 0.0064$	& $\kappa_1=1.5$	& $\omega_{1-} = 1.0$	& $\beta_1=0.0076$	\\
				$\kappa_2=1.0$	& $\omega_{2+}=5.6$	& $\beta_2= 0.5471$	& $\kappa_2=1.5$	& $\omega_{2+} = 2.5$	& $\beta_2=0.1521$ 	\\
				\hline
				\multicolumn{3}{|c||}{$\boldsymbol{\beta}_3$ \qquad} &	\multicolumn{3}{|c|}{}	\\
				\hline
				$\kappa_1=2.0$	& $\omega_{1-}=1.0$	& $\beta_1=0.0189$	& \multicolumn{3}{|c|}{}	\\
				$\kappa_2=2.0$	& $\omega_{2+}=2.3$ 	& $\beta_2=0.0732$	& \multicolumn{3}{|c|}{}	\\
				\hline
			\end{tabular}
			\caption{Examples of parameterization.}
			\label{table:dispersion_curves_annex_03a}
		\end{table}

		\begin{table}[ht]
			\centering	
			\begin{tabular}{|l|l|l||l|l|l|}
				\hline
				\multicolumn{6}{|c|}{\text{Parameters of figure \ref{fig:dispersion_curves_10}}} \\
				\hline
				\multicolumn{3}{|c||}{$\boldsymbol{\beta}_1$ \qquad} & \multicolumn{3}{c|}{$\boldsymbol{\beta}_2$ \qquad}	\\
				\hline
				$\kappa_1=\pi$	& $\omega_{1-}=1.0$	& $\beta_1= 0.0208$	& $\kappa_1=\pi$	& $\omega_{1-} = 1.0$	& $\beta_1=0.0208$	\\
				$\kappa_2=\pi$	& $\omega_{2+}=1.2$	& $\beta_2= 0.0005$	& $\kappa_2=\pi$	& $\omega_{2+} = 1.0$	& $\beta_2=-0.0039$ 	\\
				\hline
			\end{tabular}
			\caption{Examples of parameterization.}
			\label{table:dispersion_curves_annex_03b}
		\end{table}

		\begin{table}[ht]
			\centering
			\begin{tabular}{|l|l|l||l|l|l|}
				\hline
				\multicolumn{6}{|c|}{\text{Parameters of figure \ref{fig:dispersion_curves_annex_04}}} \\
				\hline
				\multicolumn{3}{|c||}{$\qquad$ \textbf{(a)} $\qquad$ } & \multicolumn{3}{c|}{$\qquad$ \textbf{(b)}$\qquad$ }	\\
				\hline
				$\kappa_1=0.5$	& $\omega_{1-}=3.5$	& $\beta_1= 0.5095$		& $\kappa_1=0.5$	& $\omega_{1+} = 3.5$	& $\beta_1=0.5497$	\\
				$\kappa_2=1.0$	& $\omega_{2-}=4.0$	& $\beta_2= 0.0018$		& $\kappa_2=0.75$	& $\omega_{2+} = 3.75$	& $\beta_2=0.1344$ 	\\
				$\kappa_3=1.5$	& $\omega_{3-}=4.5$	& $\beta_3= 3.2667$		& $\kappa_3=1.0$	& $\omega_{3+} = 4.0$	& $\beta_3=2.9091$	\\
				$\kappa_4=2.0$	& $\omega_{4-}=5.0$	& $\beta_4= 4.6104$		& $\kappa_4=1.5$	& $\omega_{4+} = 4.5$	& $\beta_4=2.7526$	\\
				$\kappa_5=2.5$	& $\omega_{5-}=5.5$	& $\beta_5= 14.6248$	& $\kappa_5=1.75$	& $\omega_{5+} = 4.75$	& $\beta_5=6.5538$	\\
				$\kappa_6=3.0$	& $\omega_{6-}=6.0$	& $\beta_6= 24.1800$	& $\kappa_6=2.0$	& $\omega_{6+} = 5.0$	& $\beta_6=8.1718$	\\
								&					&						& $\kappa_7=2.25$	& $\omega_{7+} = 5.25$	& $\beta_7=10.5234$	\\
								&					&						& $\kappa_8=2.5$	& $\omega_{8+} = 5.5$	& $\beta_8=14.8989$	\\
								&					&						& $\kappa_9=3.0$	& $\omega_{9+} = 6.0$	& $\beta_9=9.5147$	\\
								&					&						& 					& 						& $\beta_{10}=8.7373$	\\
				\hline
				\multicolumn{3}{|c||}{$\qquad$ \textbf{(c)} $\qquad$ } &  \multicolumn{3}{|c|}{}	\\
				\hline
				$\kappa_1=1.0$	& $\omega_{1+}=19.5$	& $\beta_1=0.0976$	& \multicolumn{3}{|c|}{}	\\
				$\kappa_2=1.25$	& $\omega_{2+}=19.41$ 	& $\beta_2=3.4340$	& \multicolumn{3}{|c|}{}	\\
				$\kappa_3=1.5$	& $\omega_{3+}=19.44$ 	& $\beta_3=1.2320$	& \multicolumn{3}{|c|}{}	\\
				$\kappa_4=1.75$	& $\omega_{4+}=19.62$	& $\beta_4=2.8068$	& \multicolumn{3}{|c|}{}	\\
				$\kappa_5=2.0$	& $\omega_{5+}=20$		& $\beta_5=1.1197$	& \multicolumn{3}{|c|}{}	\\
				$\kappa_6=2.25$	& $\omega_{6+}=20.63$	& $\beta_6=1.2754$	& \multicolumn{3}{|c|}{}	\\
				$\kappa_7=2.5$	& $\omega_{7+}=21.56$	& $\beta_7=0.3198$	& \multicolumn{3}{|c|}{}	\\	
				$\kappa_8=2.72$	& $\omega_{8+}=22.84$	& $\beta_8=0.2621$	& \multicolumn{3}{|c|}{}	\\	
				\hline		
			\end{tabular}
			\caption{Examples of parameterization targeting a target curve}
			\label{table:dispersion_curves_annex_04}
		\end{table}

		\begin{table}[ht]
			\centering	
			\begin{tabular}{|l|l|l|l||l|l|l|l|}
				\hline
				\multicolumn{8}{|c|}{\text{Parameters of figure \ref{fig:dispersion_curves_annex_05}}} \\
				\hline
				\multicolumn{4}{|c||}{\textbf{(a)} \qquad} & \multicolumn{4}{c|}{\textbf{(b)} \qquad}	\\
				\hline
				$\kappa_1=2.0$	& $\omega_{1-}=8.0$		& $\beta_1= 2.7090$	& 	$\beta_1= 0.9443$		&	$\kappa_1=1.0$	& $\omega_{1-} = 2.5$	& $\beta_1=2.6521$	& $\beta_1=0.9058$\\
				$\kappa_2=2.0$	& $\omega_{2+}=20.0$	& $\beta_2= 1.7230$	& 	$\beta_2= 6.6653$		&	$\kappa_2=1.0$	& $\omega_{2+} = 24.0$	& $\beta_2=1.7421$ 	& $\beta_1=8.7646$\\
				                &						&					& 							&	$\kappa_3=2.0$	& $\omega_{3-} = 8.0$	& $\beta_3=0.0281$	& $\beta_3=0.3333$\\
				                &						&					& 							&	$\kappa_4=2.0$	& $\omega_{4+} = 20.0$ 	& $\beta_4=0.1644$	& $\beta_4=-3.8671$\\				
				\hline
			\end{tabular}
			\caption{Examples of non-unique solutions.}
			\label{table:dispersion_curves_annex_05}
		\end{table}

		\begin{table}[ht]
			\centering
			\begin{tabular}{|l||l|l|l||l|l|l||l|l|l|}
				\hline
				\multicolumn{10}{|c|}{\text{Parameters of figure \ref{fig:comparaison_dispersions}}} \\
				\hline
				& \multicolumn{3}{c||}{$\qquad$ \textbf{(a)} $\qquad$} & \multicolumn{3}{c||}{$\qquad$ \textbf{(b)} $\qquad$} & \multicolumn{3}{c|}{$\qquad$ \textbf{(c)} $\qquad$} \\
				\hline
          		$\beta_p$ 	 & WN     & OP 1    & OP 2    	& WN      	& OP 1    	& OP 2    	& WN      	& OP 1    	& OP 2 	\\
				\hline
				$\beta_1$    & 31.9336 & 30.0132 & 4.0398 	& -0.4055 	&		0	&		0	& 1.3113 	& 1.2185 	& 1.2185 \\
				$\beta_2$    & 15.1452 & 12.2734 &		0	& 0.8525 	& 0.8019 	& 0.5316 	& -0.2029   & 0.0551 	& 0.0551 \\
				$\beta_3$    & 0.2799  & 	0	 & 		0	& 0.6320	& 0.4303 	& 0.7955 	& -0.1634 	&		0	&	0	 \\
				$\beta_4$    & -4.3528 & 	0	 & 6.1786 	& 0.0495 	&		0	&		0	& 0.0900 	& 0.0417 	& 0.0417 \\
				$\beta_5$    & -1.2462 & 	0	 & 13.4806 	& -0.0463 	&		0	&		0	& 0.0843 	& 0.0381 	& 0.0381 \\
				$\beta_6$    & 3.0751  & 0.2038  & 16.2458 	& 0.0446 	&		0	& 0.0704 	& 0.0067 	&		0	& 	0	 \\
				$\beta_7$    & 1.7682  & 	0	 & 11.2344 	& 0.0328 	&		0	&		0	& 0.0190 	&		0	& 	0	 \\
				$\beta_8$    & -6.0159 & 	0 	 &		0	& -0.0148 	&		0	&		0	& 0.0108 	&		0	& 	0	 \\
				$\beta_9$    & 4.8657  & 2.9468  & 8.6173  	&		0	&		0	&		0	& 0.0041 	&		0	& 	0	 \\
				$\beta_{10}$ & -1.2064 & 	0	 & 0.7174  	&		0	&		0	&		0	& 0.0057 	&		0	& 	0	 \\
				$\beta_{11}$ & -0.5215 &	0	 &		0	&		0	&		0	&		0	&		0	&		0	& 	0	 \\
				$\beta_{12}$ & 0.0741  &	0	 &		0	& 1.2250  	&		0	&		0	&		0	&		0	& 	0	 \\
				$\beta_{13}$ & 0.3604  & 	0	 &		0	&		0	&		0	&		0	&		0	&		0	& 	0	 \\
				$\beta_{14}$ & -0.0839 &	0	 &		0	&		0	&		0	&		0	&		0	&		0	& 	0	 \\
				$\beta_{15}$ & -0.0908 &	0	 &		0	&		0	&		0	&		0	&		0	&		0	& 	0	 \\
				$\beta_{16}$ & -0.1176 &	0	 &		0	&		0	&		0	&		0	&		0	&		0	& 	0	 \\
				$\beta_{17}$ & 0.2524  &	0	 &		0	&		0	&		0	&		0	&		0	&		0	& 	0	 \\
				$\beta_{18}$ & -0.1473 &	0	 &		0	&		0	&		0	&		0	&		0	&		0	& 	0	 \\
				$\beta_{19}$ & 0.0266  &	0	 &		0	&		0	&		0	&		0	&		0	&		0	& 	0	 \\
				$\beta_{20}$ & 0.0026  &	0	 &		0	&		0	&		0	&		0	&		0	&		0	& 	0	 \\
				\hline
			\end{tabular}
			\vspace{1.0em}
			\caption{Stiffness coefficients $\beta_p$ obtained for three target dispersion curves (Figures~\ref{fig:comparaison_dispersions}),  
			using three strategies: unconstrained interpolation (WN), non-negative least squares applied to the same points (OP~1),  
			and localized constrained interpolation within a selected region of interest (OP~2).}
		\end{table}

	\clearpage
	\bibliography{references}
\end{document}